\documentclass[a4paper,11pt]{article}
\pdfoutput=1 
\usepackage{jheppub} 
\usepackage[T1]{fontenc} 
\usepackage[utf8]{inputenc}

\usepackage{xcolor}
\usepackage{makecell}
\usepackage{upgreek}
\usepackage{bigints}
\usepackage{multirow}
\usepackage{booktabs}
\usepackage[normalem]{ulem}

\usepackage[T1]{fontenc} 
\usepackage[utf8]{inputenc}
\usepackage{xcolor}
\usepackage{slashed}
\usepackage{upgreek}
\usepackage{bigints}
\usepackage{multirow}
\usepackage{caption}
\usepackage{amsmath,amssymb,bbold,epsfig}
\graphicspath{{figures/}}
\usepackage{amsfonts}
\usepackage{bbm}
\usepackage{subcaption}  
\newcommand{\be}{\begin{equation}}
\newcommand{\ee}{\end{equation}}
\newcommand{\bea}{\begin{eqnarray}}
\newcommand{\eea}{\end{eqnarray}}

\definecolor{darkgreen}{HTML}{109930}

\title{\boldmath Bound-state beta decay of tritium: Path to first 
observation and novel approach to direct neutrino mass measurement}

\author[a]{Evgeny Akhmedov,}
\author[a]{Thierry Lasserre,}
\author[b]{Ferenc Gl\"{u}ck}
\author[c]{and Alejandro Saenz}

\vspace*{2.0mm}
\affiliation[a]{ Max-Planck-Institut f\"ur Kernphysik, Saupfercheckweg 1, 
69117 Heidelberg}
\affiliation[b]{Institute for Astroparticle Physics, Karlsruhe Institute of
Technology, Hermann-von-Helmholtz-Platz 1,
76344 Eggenstein-Leopoldshafen, Germany}
\affiliation[c]{AG Moderne Optik, Institut f\"{u}r Physik, 
Humboldt-Universit\"{a}t zu Berlin, Newtonstr. 15, 12489 Berlin, Germany}

\vspace*{4.0mm}
\emailAdd{akhmedov@mpi-hd.mpg.de}
\emailAdd{thierry.lasserre@mpi-hd.mpg.de}
\emailAdd{ferenc.glueck@kit.edu}
\emailAdd{Alejandro.Saenz@physik.hu-berlin.de}

\abstract
{
Bound-state $\beta$-decay of tritium, the process, in which the 
final-state electron is created in a bound atomic state of the 
produced ${\rm ^3He}$ atom instead of freely flying away, is 
predicted by the standard theory of weak interactions but has not 
been observed so far. We study the possibility of its 
experimental observation 
through the detection of photons from radiative 
decay of the excited atomic states of neutral ${\rm ^3He}$ populated by 
this process.  
We also propose a novel approach to direct neutrino mass measurement  
and sterile neutrino search based on accurate determination of the speed 
of the produced ${\rm ^3He}$ atoms through Doppler broadening of the emitted 
photon lines. 
}

\begin{document}
\maketitle
\flushbottom

\section{Introduction}
\label{sec:Introduction}

Bound-state $\beta$-decay (BSBD) is the nuclear $\beta^-$-decay process 
in which the electron is created bound to 
a vacant atomic orbital
of the final-state atom rather than in the continuum. It was first predicted 
in the late 1940s 
\cite{Daudel:1947a,sherk1949bound} and thereafter has been extensively 
studied theoretically 
(see e.g.\ \cite{Bahcall:1961zz,Takahashi:1983gsc,Budick:1983lka,
cohen1987bound,harston1993atomic,Kouzakov:2004dw,Akhmedov:2008jn, 
Faber:2009ts,McAndrew:2014iia,Gupta:2018pll,Liu:2021ptw,
Xiao:2024ldz}). BSBD was first experimentally observed 
in 1992 in fully ionized ${\rm ^{163}Dy}$ \cite{Jung:1992pw}, and has 
subsequently been detected in three additional fully 
ionized atomic species -- ${\rm ^{187}Re}$, ${\rm ^{207}Tl}$ and 
${\rm ^{205}Tl}$ \cite{Bosch:1996zz,Ohtsubo:2005zz,E121:2024zpp}. 
The standard weak interaction theory predicts that tritium should 
experience BSBD as well, with relatively large branching ratio with 
respect to the usual, continuum-state $\beta$-decay (CSBD): 
$\Gamma_{\rm BSBD}/\Gamma_{\rm CSBD} = {\cal O}(1\%)$.  
However, this process has not been observed so far. 

BSBD of atomic tritium, 
\be
 {\rm T}\to
{\rm ^3He}+\,\bar{\nu}_e\,,
\label{eq:T1}
\ee
creates a neutral ${\rm ^3He}$ atom and an electron antineutrino in the final 
state. As the $\bar{\nu}_e$ 
is practically unobservable, the process can only be observed through 
the detection of ${\rm ^3He}$. In this paper we suggest a way of doing this. 
Calculations show that a significant 
fraction of ${\rm ^3He}$ atoms are produced in several low-lying 
excited states. Their subsequent radiative decays produce  
characteristic atomic photons
that can be readily detected, providing a clear signature of tritium BSBD. 

More importantly, this opens up a novel possibility 
of direct neutrino mass measurement. Two-body nature of the final state 
of process~(\ref{eq:T1}) implies that, for a fixed neutrino mass, the 
produced ${\rm ^3He}$ atoms and $\bar{\nu}_e$ are monoenergetic. 
The velocities of ${\rm ^3He}$ exhibit a weak dependence on neutrino mass; 
therefore, accurate measurements of these velocities would determine  
neutrino masses.   

How can the velocities of the produced ${\rm ^3He}$ be determined? 
Time-of-flight techniques cannot be used in this case, as the instant of 
time when the decay of the parent tritium occurred is unknown, and there is 
no electron emission that could be used as a time tag. However, the 
velocities of ${\rm ^3He}$ atoms produced in excited atomic states can 
be determined through Doppler broadening of the lines of the emitted 
photons. Thus, the same radiative decay processes that can be used for the 
observation of BSBD of tritium can also be employed for probing neutrino 
masses. The requirements for these two types of experiments are, however, 
different. While for observation of tritium BSBD it is sufficient to 
detect the characteristic ${\rm ^3He}$ de-excitation photons  
(provided that backgrounds are under control), 
in order to probe neutrino masses one has to measure very accurately 
the lineshapes of the emitted photons.  This is much more demanding. 
The current state-of-the-art in quantum optics allows atomic 
frequency measurements with accuracies reaching   
$\Delta\omega/\omega_0\sim 10^{-18}$ \cite{Kawasaki:2024pqt,Marshall:2025ahl}. 
The accuracies of lineshape measurements are more modest, currently in 
the $10^{-4}-10^{-11}$ range 
\cite{Bekker:2019ons,fredrick2022thermal,Thorlabs:01}, 
but progress in this field is very fast. In addition to very accurate 
lineshape measurements, cryogenic temperatures will be necessary in 
order to suppress thermal motion of the parent tritium atoms. 

Since electron antineutrinos produced in BSBD are linear superpositions 
of different neutrino mass eigenstates, the 
lineshapes of the individual photon lines should exhibit characteristic 
kinks, similarly to the kinks in the electron spectra sought in CSBD 
experiments \cite{Formaggio:2021nfz,KATRIN:2024cdt,PTOLEMY:2019hkd}. 
There are, however, important differences. While direct neutrino mass 
measurement experiments based on nuclear $\beta$-decay usually measure 
electron energies near the endpoint of the spectrum, our proposal relies 
on photon frequency measurements. In this respect it is similar to 
approaches based on cyclotron radiation emission spectroscopy (CRES), such 
as Project 8 \cite{Project8:2017nal} or QTNM \cite{Amad:2024jod}. 
What sets our approach apart from all CSBD-based neutrino mass 
measurements is the scaling of the relative number of events in the 
neutrino-mass-sensitive region: this fraction scales as $(m_\nu/Q)^2$, 
whereas in CSBD-based experiments it scales as $(m_\nu/Q)^3$ (where 
$Q$ is the $Q$-value of the process). 
Thus, the number of ``useful'' events is increased by a very large factor 
of $Q/m_\nu$, which is a direct consequence of 
the two-body decay kinematics in BSBD, as opposed to the three-body 
kinematics of the ordinary $\beta$-decay. 

Direct neutrino mass measurement experiments based on orbital electron 
capture (EC) \cite{Gastaldo:2013wha,Alpert:2014lfa,Martoff:2021vxp} 
also involve 
two-body final states. However, EC produces 
highly excited daughter atoms that de-excite via  
$X$-ray cascades and Auger electrons, requiring calorimetric measurements 
for precise determination of the released energy. In contrast, 
BSBD leaves the final-state 
atoms in the ground state or in low-lying excited states; the latter 
de-excite by emitting optical or extreme ultraviolet (XUV) photons, 
which are much easier to detect. 

In this paper we focus primarily on the conceptual issues 
and general aspects of the proposed observation of BSBD of tritium 
via detection of photons coming from de-excitation of the final-state 
${\rm ^3He}$ atoms, as well as of the novel approach to direct neutrino 
mass measurement that it offers. We do not discuss specific experimental 
setups or the associated systematic uncertainties, as these lie outside 
the scope of the present work and will be addressed in future studies.  

The units $\hbar=c=1$ are used throughout the paper. 

\section{
\label{sec:observ}Tritium BSBD observation}

\subsection{\label{sec:rates}$\rm ^3He$ atomic states and radiative 
transitions relevant for BSBD}

The rate of BSBD of atomic tritium has been evaluated by a number of 
authors. To our knowledge, the most accurate calculations were carried out 
in Ref.~\cite{harston1993atomic}, where the BSBD branching ratio 
was found to be $\Gamma_{\rm BSBD}/\Gamma_{\rm CSBD} = 0.55\%$. The 
authors also calculated separately the fractions of BSBD transitions to 
the ground state and to a number of excited states of the final-state 
neutral ${\rm ^3He}$ atoms. In 
Table~\ref{tab:fractions} we present these fractions for the ground state 
as well as for $1s2s$ and $1s3s$ excited states of ${\rm ^3He}$ 
(combined singlet and triplet), which together constitute over 95\% of all 
BSBD transitions. The branching fractions of the excited states should be 
distributed between singlet and triplet ${\rm ^3He}$ statistically, 
with the ratio 1:3. 

\begin{table*}[h]
\centering
\begin{tabular}{|c|c|}
\hline
\hline
\bf{Final atomic state of
$\rm{^3He}$} & \bf{Branching fraction} \\
 \hline
$1s^2$ ground state (singlet) & 57\% \\
\hline
$1s2s$ states (singlet+triplet) & 32.5\% \\
$1s3s$ states (singlet+triplet) & 5.6\% \\
\hline
\end{tabular}
\caption{\small{Fractions of tritium BSBD transitions to the ground 
state and a few excited states of ${\rm ^3He}$ \cite{harston1993atomic}.}}
\label{tab:fractions}
\end{table*}

\begin{figure}[!ht]
\begin{center}
\vspace*{-2mm}
\colorbox{white}{\includegraphics[width=15.2cm]{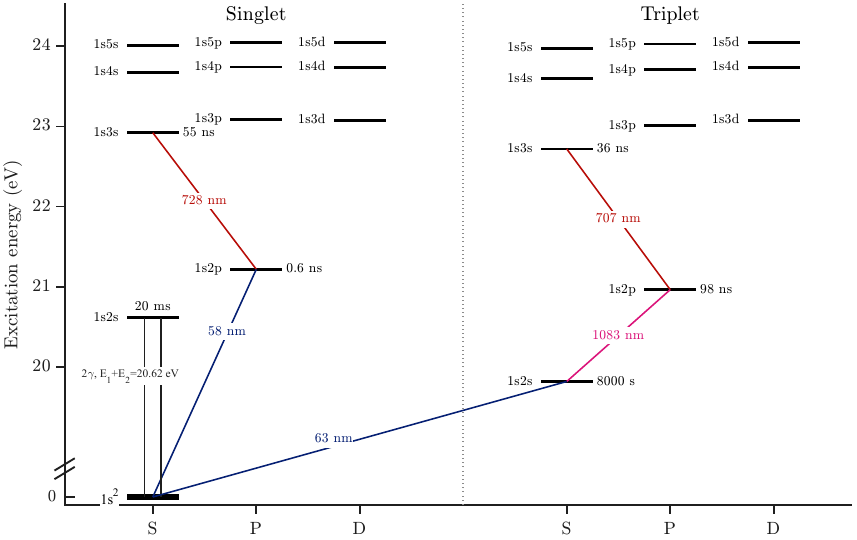}}
\vspace*{-5mm}
\end{center}
\caption{\small Partial atomic level schemes for singlet 
(left) and triplet (right) ${\rm ^3He}$ atoms. Atomic transitions of interest 
are shown by colored lines. The data are taken from \cite{NIST_ASD}.} 
\label{fig:levels}
\end{figure}

\noindent
The level schemes of singlet and triplet ${\rm ^3He}$ atoms as well as 
radiative decays of interest are shown in Figure~\ref{fig:levels}.
We summarize the properties of the relevant states and transitions 
in Table~\ref{tab:features}. 
The options for the observation of BSBD of tritium include:

\vspace*{1.2mm}
I. {\em Detection of photons from radiative decays of singlet or 
triplet $1s2s$ states.} 

\vspace*{1mm}
\noindent
{\em (a)} The singlet $1s2s$ state decays to the $1s^2$ ground state by 
emitting a pair of E1 photons (single photon emission is highly  
suppressed). Coincidence detection of photon pairs whose energies satisfy  
$\omega_1+\omega_2=E_{\rm exc.}=20.62$\;eV could serve as a signature 
of BSBD. 

There are, however, certain limitations related to this channel. 
At least one of the two photons in the emitted pair  
is in the XUV range, 
typically both. Detecting such photons is more difficult than detecting 
e.g.\ optical photons; in addition, energy resolution of XUV photon 
detectors is at best marginal. Up to now, radiative decays of the singlet 
$1s2s$ states of neutral helium atoms (either ${\rm ^3He}$ or 
${\rm ^4He}$) have not been directly observed. 
Their lifetime, $\tau\simeq 20$\;ms, has only been measured 
indirectly through time-of-flight techniques. In addition to the already 
mentioned poor energy resolution of photodetectors in the XUV region, 
the observation is complicated by the fact that 
coincident detection requires strong sources (sufficient
amount of excited helium is needed). 
This, however, does not 
seem to constitute a problem in tritium $\beta$-decay experiments 
such as KATRIN \cite{KATRIN:2024cdt}: 
BSBD of tritium is expected to produce 
very large amounts of singlet ${\rm ^3He}$ atoms in the metastable $1s2s$ 
state. Moreover, lack of energy resolution of XUV photodetectors may
to some extent be mitigated by coincidence detection provided that a good 
time resolution can be achieved.
{\small{
\begin{table*}[t]
\centering
\begin{tabular}{|c|c|c|c|c|c|}
\hline
\hline
\bf{State}
& $\mathbf{E_{\rm exc.}}\;{\rm(eV)}$ & 
$\boldsymbol \tau$ 
& {\bf Decay mode} & 
$\mathbf{E_{decay}}$\;(eV) &{\bf Wavelength}\\
\hline
singlet $1s2s$ & 20.62 & 20\,ms & 
\makecell{
$1s2s\to 1s^2$(g.s.), 
2$\gamma$, E1,\!\\ $\omega_1+\omega_2=E_{\rm exc.}$ }
& 20.62 & continuum \\
triplet $1s2s$ & 19.82 & 7870\;s & $1s2s\to 1s^2$(g.s.), M1 & 19.82 & 
62.56\;nm (XUV)\\
singlet $1s3s$ & 22.93 & 54.64\;ns & $1s3s\to 1s2p$(singl.), E1 & 1.703 & 
728.1\;nm (Red)\\
triplet $1s3s$ & 22.72 & 35.92\;ns & $1s3s\to 1s2p$(tripl.), E1 & 1.755 & 
706.5\;nm (Red)\\
singlet $1s2p$ & 21.23 & 0.56\;ns & $1s2p\to 1s^2$(g.s.), E1 & 21.23 & 
58.4\;nm (XUV)\\
triplet $1s2p$ & 20.96 & 98\;ns & $1s2p\to 1s2s$(tripl.), E1 & 1.145 & 
1083\;nm (NIR)\\
\hline
\end{tabular}
\caption{\small{Properties of the relevant ${\rm ^3He}$ excited states and
their radiative de-excitation transitions \cite{NIST_ASD}. The table lists
the excitation energy $E_{\rm exc.}$, lifetime $\tau$, decay mode, transition
type, radiative decay energy $E_{\rm decay}$, and photon wavelength for each
state.}}
\label{tab:features}
\end{table*}
Thus, if BSBD is identified via this channel, this will not only be an 
observation of tritium BSBD, but also the first direct observation of
2-photon decay of the singlet $1s2s$ metastable state or helium. 

\noindent  
{\em (b)} The triplet $1s2s$ excited states of ${\rm ^3He}$ decay to the 
$1s^2$ ground state by emitting a single M1 photon with energy 
$\omega=19.82$\;eV (wavelength $\lambda=62.56$\;nm, in the XUV range). 
The observation of this transition faces the same difficulties with 
XUV photon detection as the decay of the singlet $1s2s$ state.
In addition, the triplet $1s2s$ state is extremely long-lived 
($\tau=7870$\;s), which makes it particularly vulnerable to collisional 
de-excitation and thus very difficult to observe via radiative decay. 

\vspace*{2mm}
II. {\em Detection of photons from radiative decays of singlet or triplet 
$1s3s$ states to the corresponding 
$1s2p$ states.} 

\vspace*{1mm}
These transitions are very similar in singlet and triplet ${\rm ^3He}$. 
The singlet $1s3s$ state (lifetime $\tau=54.64$\;ns) decays to the singlet 
$1s2p$ state by emitting an E1 photon with energy 
$\omega=1.703$\;eV (wavelength $\lambda=728.1$\;nm). 
The triplet $1s3s$ state (lifetime $\tau=35.92$\;ns) decays to the 
triplet $1s2p$ state by emitting an E1 photon with energy 
$\omega=1.755$\;eV (wavelength $\lambda=706.5$\;nm).%
\footnote{In triplet helium, $1s2p$ actually denotes 
a triplet of fine-structure states with angular momenta $J=0,1,2$ 
(denoted $1s2p\,^3\!P_{0,1,2}$), separated by very small 
energy intervals, $\sim 10^{-4}$\;eV. 
The $1s3s\to 1s2p$ transitions thus produce a triplet of nearly degenerate 
red lines. However, these can currently only be resolved by 
high-resolution laser spectroscopy, so their splitting is irrelevant for 
BSBD observation. The same holds for near-infrared $1s2p\to 1s2s$ 
transitions in triplet helium discussed in point III(b) below. 
Fine-structure splitting of triplet helium lines should, however, 
be taken into account in probing neutrino masses via Doppler 
emission spectroscopy as discussed in section \ref{sec:mass}. 
It will 
will not hinder the neutrino mass measurements.} 
Both transitions correspond to the well-known red lines of neutral helium, 
which are routinely observed in atomic spectroscopy and are often used for 
plasma diagnostics and in astrophysical studies. 

\vspace*{2.0mm}
III. {\em Detection of photons from radiative decays of singlet or triplet 
$1s2p$ states populated by the preceding decays of the corresponding   
$1s3s$ states.} 

\vspace*{1.5mm}
\noindent
{\em (a)} The singlet $1s2p$ state decays almost exclusively (with 
branching ratio 99.9\%) to the ground state of ${\rm ^3He}$ by 
emitting an E1 photon with energy $\omega=21.23$\;eV (wavelength 
$\lambda=58.4$\;nm, in the XUV range). The lifetime of this state is very 
short, $\tau=0.56$\;ns. 

\vspace*{1.5mm}
\noindent
{\em (b)} The triplet $1s2p$ state decays to the $1s2s$ state of triplet 
helium by emitting an E1 photon with energy $\omega=1.145$\;eV (wavelength 
$\lambda=1083$\;nm). 
The lifetime of the initial 
state is $\tau=98$\;ns. 
This transition corresponds to the well-known near-infrared (NIR) line 
in the helium spectrum which is widely used for precision tests of 
quantum electrodynamics, laser cooling and trapping of helium, and in 
astrophysical studies.  

\vspace*{2mm}
IV. {\em Detection of sequences of radiative transitions in prompt or delayed 
coincidence.}
 
\vspace*{1.5mm}
\noindent
To reject possible accidental backgrounds, coincidence detection of 
photons emitted in the following sequences of transitions may also 
be helpful: 

\vspace*{1.5mm} 
1. Red and NIR photons from the transitions in triplet 
${\rm ^3He}$:

\be
1s3s ~~\overset{\rm 706.5\,nm}
{-\!\!\!-\!\!\!-\!\!\!\longrightarrow}~~
1s2p
~~\overset{\rm 1083\,nm}
{-\!\!\!-\!\!\!-\!\!\!\longrightarrow}~~ 1s2s\qquad
\label{eq:seq1}
\vspace*{1.5mm}
\ee
in delayed coincidence (lifetime of the $1s2p$ state ${\rm \sim 100\,ns}$).

\vspace*{1.5mm}
2. Red and XUV photons from the transitions in singlet 
${\rm ^3He}$: 
\vspace*{1.5mm}
\be
1s3s ~~\overset{\rm 728.1\,nm} 
{-\!\!\!-\!\!\!-\!\!\!\longrightarrow}~~1s2p
~~\overset{\rm 58.4\,nm}
{-\!\!\!-\!\!\!-\!\!\!\longrightarrow}~~ 1s^2\qquad
\label{eq:seq2}
\vspace*{0.5mm}
\ee
in prompt coincidence (lifetime of the $1s2p$ state ${\rm \sim 0.6\,ns}$).

\vspace*{2mm}
Summarizing, the most promising approaches for observing 
BSBD of tritium appear to be 
\begin{itemize}
\item 
Detection of red-line photons from the $1s3s\to 1s2p$ transitions in 
either triplet or singlet ${\rm ^3He}$ (or both)

\item 
Detection of NIR photons from the $1s2p\to 1s2s$ 
transitions in triplet ${\rm ^3He}$ 

\item 
Delayed coincidence detection of red and NIR photons in triplet 
${\rm ^3He}$ [eq.~(\ref{eq:seq1})] 

\item 
Prompt coincidence detection of red and 
XUV photons in singlet ${\rm ^3He}$ [eq.~(\ref{eq:seq2})] 
 
\end{itemize}
These transitions are shown in Fig.~\ref{fig:levels} by colored lines.

\subsection{\label{sec:eventRates} Radiative and collisional 
de-excitation of ${\rm ^3He}$. Event rates}
We now estimate the expected numbers of events for the proposed 
observation of BSBD of tritium. 
We assume an atomic tritium source with activity 
$10^8$\;Bq, corresponding to  $N_0=5.6\times 10^{16}$ tritium atoms 
($10^{-3}$ of the amount used in the KATRIN experiment).%
\footnote{The cases of molecular and mixed atomic/molecular 
tritium sources are discussed in section~\ref{sec:irred} and 
Appendix~\ref{sec:appMixed}.}

{} From the tritium CSBD rate $\Gamma_1={\rm(17.75\;yr)^{-1}=
1.786\times 10^{-9}\,s}^{-1}$ and the BSBD branching ratio of 0.55\%, 
one finds the total BSBD rate  
$\Gamma_{\rm BSBD}=9.89\times 10^{-12}~{\rm s}^{-1}$. 

We denote by $\Gamma_2$ the BSBD production rate of ${\rm ^3He}$ atoms 
in a specific excited atomic state under study. For definiteness, 
we focus below on the 
production of 
${\rm ^3He}$ in the triplet $1s3s$ excited state. 
{}From Table~\ref{tab:fractions} we find 
\be
\Gamma_2=
4.15\times 10^{-13}~{\rm s}^{-1}.
\label{eq:Gamma2}
\ee
The production rates for the other excited states of ${\rm ^3He}$ 
atoms discussed above can also be readily obtained from 
Table~\ref{tab:fractions}. 

Since the radiative decays of all excited ${\rm ^3He}$ states 
(including the metastable ones) are much faster 
than BSBD of tritium, the emission rate of ${\rm ^3He}$ de-excitation 
photons  is dominated by BSBD. 
From eq.~(\ref{eq:Gamma2}) we then obtain the photon production rate 
\be
\frac{d{N}_{\rm ph}}{dt}
=N_0\Gamma_2\simeq 
23000\;{\rm s}^{-1}.
\ee
Thus, the number of events accumulated over 24 hours is $\simeq 2\times 
10^9$. The number of actually observed events will of course be smaller, 
since the incomplete coverage of the inner walls of the experimental 
cavity by photodetectors and the detection efficiency must be taken into 
account. Nevertheless, the expected event statistics remains very high. 

In estimating the event numbers, we assumed that all excited ${\rm ^3He}$ 
atoms de-excite radiatively, i.e., we ignored possible collisional 
de-excitation. Let us now estimate its possible effects.  
Non-radiative relaxation of the excited states of ${\rm ^3He}$ atoms can 
occur through their collisions with either the inner walls of the 
experimental cavity or with other particles present in it, primarily 
undecayed tritium atoms. The velocity of the produced ${\rm ^3He}$ 
atoms is $v_0\simeq 2$\;km/s (see section~\ref{sec:speed}). Assuming an 
experimental cell of linear size $L\sim 10$\;cm,  
the average time for an excited $^3$He atom to hit an inner wall
and de-excite is 
\be
t_{c1}\simeq \frac{L}{v_0}\simeq 50\,\mu{\rm s}\,.
\vspace*{-0.1cm}
\ee
Next, we estimate 
the characteristic time of ${\rm ^3He}$ de-excitation via collisions with 
tritium atoms. The tritium number density in a volume of 
$({\rm 10\,cm})^3$ is  
\vspace*{-1mm}
\be
n_T=\frac{5.6\times 10^{16}}{10^3\;{\rm cm^3}}=
5.6\times 10^{13}{\rm cm^{-3}}.
\ee
Assuming a collisional de-excitation cross section 
\,$\sigma_c\simeq 10^{10} \,b\,=10^{-14}\,{\rm{cm^2}}$ (a conservative 
estimate), one finds 
\[
t_{c2}=(n_T \sigma_c v_0)^{-1}\simeq 10\;\mu{\rm s}.
\vspace*{1mm}
\]
Comparing these de-excitation times with the radiative lifetimes of the 
excited ${\rm ^3He}$ states given in Table~\ref{tab:features}, one 
sees that, under the discussed conditions, the collisional 
de-excitation processes are much slower than radiative decay -- and 
therefore can be safely neglected -- for all excited helium states 
except the metastable $1s2s$ ones. For the singlet metastable state, 
BSBD observation might still be possible if detectors 
of significantly larger sizes are used; 
however, for the 
extremely long-lived triplet $1s2s$ state, collisional relaxation 
will likely dominate under any realistic experimental conditions.

\section{Backgrounds}
\label{sec:bckgr}

There are two types of possible backgrounds to BSBD of tritium 
that can hinder its unambiguous experimental observation:

\begin{itemize}
\item
Backgrounds from processes different from BSBD but leading to the same
excited states of neutral ${\rm ^3He}$ atoms as those we are
interested in.
These are irreducible backgrounds that can precisely mimic BSBD. 

\vspace*{4mm}
\item
Accidental backgrounds from other processes that could lead to similar 
signals (e.g.\ photons from bremsstrahlung in CSBD) in the case of 
insufficient detector energy or time resolution. 
\end{itemize}

\subsection{\label{sec:irred}CSBD and irreducible backgrounds}

\vspace*{2mm}
\noindent
{\em (i) Production of neutral ${\rm ^3He}$ in CSBD of \,${\rm T_2}$.}

\vspace*{1mm}
\noindent
Consider now the case of a molecular tritium source and the 
associated backgrounds to BSBD. One possible 
background is related to 
CSBD of molecular tritium,  
\be
{\rm T_2\,\to ~^3He T}^++e^-+\bar{\nu}_e\,. 
\label{eq:CSBDmolec}
\ee
It produces molecular ions ${\rm ^3He T}^+$, 
about half  
of which dissociate 
\cite{Jonsell:1999aq,Saenz:2000dul,Bodine:2015sma,TRIMS:2020nsv}. The 
main dissociation channels are (ref.~\cite{Bodine:2015sma}, table V):
\be
{\rm ^3He T^+\!\to T^++\,^3He}\qquad (\sim66\%)\,,\;
\label{eq:Chan1}
\ee
\be
{\rm ^3He T^+\!\to T\,+\,^3He^+}\qquad (\sim 18\%)\,, 
\label{eq:Chan2}
\ee
where the dissociation fractions are shown in the parentheses.%
\footnote{The quoted dissociation fractions were obtained in 
the sudden approximation.}\,%
Other channels, such as e.g.\ double ionization of the produced 
${\rm ^3He}$, are not relevant to our discussion. 

Channel (\ref{eq:Chan1}) may populate, among others, the $1s2s$ or 
$1s3s$ excited states of neutral ${\rm ^3He}$, if the molecular ions 
${\rm ^3He T^+}$ are produced in sufficiently high excited electronic 
states. This would create an irreducible background for the 
tritium BSBD observation method proposed here. 
Unfortunately, a reliable evaluation of the fraction of CSBD transitions 
in T$_2$ leading to such excited states is quite involved and is not 
currently available. Given that the tritium CSBD rate $\Gamma_1$ is 
three to four orders of magnitude larger than the partial BSBD rates 
$\Gamma_2$ of interest, even a modest fraction of $1s2s$ or $1s3s$ 
excited neutral $^3$He atoms produced via channel (\ref{eq:Chan1}) could 
make the observation of tritium BSBD using the approach suggested here 
essentially impossible in molecular tritium. In that case, one would 
need to search for tritium BSBD using atomic tritium sources. 

Interestingly, the fraction of the “unwanted” CSBD processes  
leading to the excited $^3$He state under study  
could, in principle, be probed experimentally with 
mixed atomic/molecular tritium sources, provided that the 
atomic-to-molecular ratio can be varied.%
\footnote{We are grateful to Sergey Vasiliev for this observation.}
We concentrate here on the $1s3s$ states of singlet and triplet helium, 
for which collisional de-excitation can be neglected and the numbers of 
the emitted photons essentially coincide with the numbers of atoms 
produced in the corresponding excited states, as discussed in 
section~\ref{sec:eventRates}. 
Assume that the initial total number of tritium atoms in the mixture is 
$N_0$, and that the atomic fraction is $k$. The mixture then contains 
$N_a=N_0 k$ tritium atoms and $N_m=\frac{1}{2}N_0(1-k)$ 
molecules. We denote by $r$ the fraction of the CSBD transitions in 
molecular tritium that lead to the excited $1s3s$ states of neutral 
helium.  The rate of these ``unwanted'' transitions 
is then $\Gamma_1 r$, and the number of events accumulated over a time 
interval $t$ is 
\be
2N_m \Gamma_1 r t=N_0 \Gamma_1 r (1-k) t\,.
\label{eq:CSBDcontrib}
\ee
We also need both the atomic and molecular rates of BSBD transitions 
to the excited atomic state under study. As discussed in 
section~\ref{sec:rates}, the atomic rates $\Gamma_2$ are accurately 
known from the calculations; unfortunately, no calculations of BSBD 
rates in molecular tritium are available. We therefore parameterize the 
molecular BSBD rate (per constituent atom) for the transition with 
excitation of the relevant ${\rm ^3He}$ state as $(1-\delta)\Gamma_2$. 
The molecular BSBD rates are not expected to be drastically different 
from the corresponding atomic rates, i.e.\ one expects 
$1-\delta={\cal O}(1)$. 
The total number of the atomic and molecular BSBD events accumulated 
over the time interval $t$ is $N_0 t [\Gamma_2k+(1-\delta)\Gamma_2(1-k)]$. 
Combining this with the CSBD contribution (\ref{eq:CSBDcontrib}) 
and taking into account that the number of de-excitation photons 
coincides with that of the excited helium states, 
we find 
\be
N_{\rm ph}=N_0 t\Big[\Big(r -\frac{\Gamma_2}{\Gamma_1}\delta\Big)
\Gamma_1(1-k)+\Gamma_2\Big]\,.
\label{eq:Nph1}
\ee   
Because $\Gamma_2/\Gamma_1\sim 10^{-4}$ and $\delta={\cal O}(1)$ 
while $r$ may be as large as a few percent \cite{Bodine:2015sma}, 
one can to a first approximation neglect the term $(\Gamma_2/\Gamma_1)
\delta$ in eq.~(\ref{eq:Nph1}), which yields 
\be
N_{\rm ph}\simeq N_0 t[\Gamma_1 r (1-k)+\Gamma_2]\,.
\label{eq:Nph2}
\ee   
Thus, by varying the atomic fraction $k$ and studying the dependence 
of the number of emitted photons on $1-k$, one could in 
principle determine the value of $r$,  
which governs the irreducible background from CSBD of tritium. 
If the $(1-k)$-independent term in (\ref{eq:Nph2}) can be reliably 
shown to be nonzero, this would constitute an unambiguous observation 
of tritium BSBD and the first experimental measurement of its 
partial rate $\Gamma_2$. See Appendix~\ref{sec:appMixed} for details.

It should be stressed that the above discussion of BSBD and CSBD 
for mixed molecular/atomic tritium sources is necessarily 
simplified and schematic. It does not take into account important 
experiment-specific effects, such as adsorption of molecular 
tritium on the inner walls of the experimental cavity or possible 
time dependence of the atomic fraction $k$. These features depend 
on the particular experimental setup and cannot be assessed in  
an experiment-independent way.

\vspace*{2mm}
\noindent
{\em (ii) ${\rm ^3He}^+-e^-$ recombination.}

According to ref.~\cite{Bodine:2015sma}, about 10\% of all CSBD 
transitions in molecular tritium produce singly charged helium ions 
${\rm ^3He}^+$ through the dissociation channel (\ref{eq:Chan2}) 
of ${\rm ^3HeT}^+$ molecular ions.%
\footnote{This follows from eq.~(\ref{eq:Chan2}), taking 
into account that only about 50\% of the produced molecular ions 
${\rm ^3He T}^+$ dissociate.} 
In the case of CSBD of atomic tritium, 
${\rm ^3He}^+$ ions are produced in 100\% of the decays. Radiative 
recombination of these ions with free electrons present inside the 
experimental cavity produces neutral helium atoms. This process may 
populate, among others, the excited atomic states of ${\rm ^3He}$ 
of interest here, and therefore can be another source of irreducible 
background to BSBD of tritium. The total cross section of radiative 
recombination is, however, rather small:  
$\sigma_{rr}={\rm 7.6\times 10^{-18}\,cm^2}$ in the limit of vanishing 
electron velocity $v_e$, and it decreases rapidly with 
increasing $v_e$ (see, e.g., \cite{kotelnikov2019electron}). The velocity 
distribution of electrons is experiment-dependent, and therefore 
the recombination rate cannot be assessed in a general way. 
If the background from ${\rm ^3He}^+-e^-$ recombination in a particular  
experiment is found to be of concern, electrons and ${\rm ^3He}^+$ ions 
can be efficiently removed by applying suitably configured transient   
electric and magnetic fields.

\subsection{\label{sec:accid}Accidental backgrounds and their rejection}

One possible source of accidental backgrounds in tritium BSBD experiments 
is inner bremsstrahlung accompanying CSBD of tritium. This process 
produces photons with energies ranging from zero up to 18.6\,keV, 
a small fraction of which have energies overlapping with those of 
the atomic transitions in helium of interest here.  

As an example, consider the red-line photons emitted in the radiative 
decays of the $1s3s$ states of singlet or triplet helium, discussed in 
section~\ref{sec:rates} (with energies 1.703\,eV and 1.755\,eV, 
respectively). Numerical calculations based on the standard theory of 
inner bremsstrahlung \cite{Gluck:1997km} show that the fractions of 
tritium CSBD events producing photons in the energy 
interval ${\rm [1.5\,eV,\, 2\,eV]}$ is approximately $10^{-5}$, which 
is about a factor of 20 smaller than the BSBD branching ratios 
for the excitation of the $1s3s$ states of ${\rm ^3He}$. This background 
can be further suppressed by using optical filters to reduce the spectral 
window for photon detection. To a good accuracy, it can be neglected; since 
it is small and accurately known, it can also be simply subtracted.

In general, accidental backgrounds of various origins can be suppressed  
by using narrow-bandpass optical filters. Accidentals can also be 
efficiently rejected by detecting prompt or delayed coincidences 
between the photons from the cascade decays $1s3s\to 1s2p\to 1s^2$ 
in singlet helium or $1s3s\to 1s2p\to 1s2s$ in triplet helium, as 
discussed in section~\ref{sec:rates}. 

\section{
\label{sec:mass}
Direct neutrino mass measurements through ${\rm ^3He}$ speed 
determination}

We now consider the possibility of determining neutrino mass 
by measuring the velocities of ${\rm ^3He}$ atoms produced in tritium 
BSBD. In what follows, we restrict ourselves to BSBD of atomic tritium.

\subsection{\label{sec:speed}Neutrino mass and ${\rm ^3He}$ velocities}

For a given neutrino mass, the 
velocity of a neutral ${\rm ^3He}$ produced in tritium BSBD 
has a fixed value in the tritium's rest frame. It is defined by the 
kinematics of two-body decay, and in the limit of vanishing neutrino 
mass is given by  
\be
v_0=\frac{Q}{M}\simeq 1.986\,{\rm km/s}\,.
\label{eq:v0}
\ee
Here $M\equiv(M_i^2+M_f^2)/(M_i+M_f)$, where 
$M_i$ and $M_f$ are the atomic masses of tritium and ${\rm ^3He}$. 
The $Q$-value of the process is the difference of these masses: 
$Q=M_i-M_f\simeq 18.6$\,keV.  
For transitions to an excited atomic state of ${\rm ^3He}$ with the 
excitation energy $E_{\rm exc}$, the quantity $M_f$ must be replaced by 
$M({\rm ^3He})+E_{\rm exc}$, and the $Q$-value of the process must be 
modified accordingly. For tritium BSBD with the production of neutrino 
mass eigenstate $\nu_k$ of mass $m_k$, the velocity of the final-state  
helium atom is 
 \be
v_k\simeq\frac{\sqrt{Q^2-m_k^2}}{M}\approx v_0\Big(1-\frac{m_k^2}{2Q^2}\Big)\,.
\label{eq:v}
\ee
The first approximate equality here holds to an accuracy of about 
$10^{-11}$; the exact relation is given in Appendix~\ref{sec:appA}. 
The velocity change of the produced helium atom due 
to nonzero neutrino mass is 
\be
\Delta v\equiv v_0-v_k\simeq \frac{m_k^2}{2MQ}\simeq 
2.87\times 10^{-4}\,(m_k/{\rm eV)^2\,cm/s}\,.
\label{eq:Deltav}
\ee
Thus, the velocity of the ${\rm ^3He}$ 
atom is reduced relative to $v_0$ by about 3\,m/s for a 
keV-scale (sterile) neutrino, but only by $\sim$7\,nm/s for $m_k\simeq 
50$\,meV, which corresponds to the mass of the heaviest active neutrino.

\subsection{\label{sec:doppler}\!Neutrino mass measurement through 
the Doppler effect}
The idea of using the Doppler effect to observe tritium BSBD 
and probe neutrino masses was first proposed in 
ref.~\cite{cohen1987bound}. The authors suggested using  
tunable lasers to excite the metastable singlet helium (produced in  
tritium BSBD with a sizable branching ratio) from the $1s2s$ state 
to the $1s3p$ state and to detect the subsequent  
de-excitation photons. 
The laser can be tuned to this transition approximately; if the 
detuning is sufficiently small, it can be compensated by the 
Doppler shift arising from the nonzero velocity of the excited 
helium atoms. By scanning the detuning, one can determine this 
velocity. The method is thus based on laser-induced fluorescence 
and Doppler absorption spectroscopy. 

The advantage of this approach is that the resonant spectroscopy allows 
photon frequency measurements with extremely high resolution. The method, 
however, has some limitations: small irradiation areas  
and additional line broadening sources, such as power broadening and transit 
time broadening. 

In contrast, our approach relies on Doppler emission spectroscopy: 
we propose to accurately measure the profiles of the emission lines of 
the excited helium atoms corresponding to the transitions discussed in 
section~\ref{sec:rates}. The relevant excited states of ${\rm ^3He}$ 
are populated by tritium BSBD itself, and no laser-induced optical 
pumping is required.

We first consider the case when the velocities of the parent tritium 
atoms are negligibly small. The effects of thermal motion of 
tritium will be discussed in section~\ref{sec:thermal}.    
 
The profile of a photon line emitted by an isolated excited atom at rest 
has the Lorentzian form:
\be
L(\omega)=\frac{\Gamma/2\pi}{(\omega-\omega_0)^2+\Gamma^2/4}\,.
\label{eq:Lorentz}
\ee
Here $\omega_0$ is the central frequency of the line and $\Gamma$ is 
the sum of the natural linewidths of the initial and final atomic 
states. 

If the parent atom moves with speed $v$, the lineshape 
is modified by the Doppler effect. To linear order in $v/c$, one finds   
\be
L_{v,\cos\theta}(\omega)=\frac{\Gamma/2\pi}{(\omega-\omega_0-
\omega_0\frac{v}{c}\cos\theta)^2+\Gamma^2/4}\,.
\label{eq:Doppl1}
\ee
The Doppler shift $\omega_0\frac{v}{c}\cos\theta$ depends on the angle 
$\theta$ between the velocity of the excited helium atom and the 
direction of photon emission. Since BSBD of tritium produces helium 
atoms with isotropically distributed velocities, one must average the 
lineshape (\ref{eq:Doppl1}) over $\cos\theta$, which yields   
\be
F(\Delta;a,\Gamma)\equiv\frac{1}{2}\int_{-1}^1 d\!\cos\theta
L_{v,\cos\theta}(\omega)=\frac{1}{2\pi a}\left\{
\arctan\Big[\frac{2(\Delta+a)}{\Gamma}\Big]-\arctan\Big[\frac{2(\Delta-a)}
{\Gamma}\Big]
\right\}.
\label{eq:Dopp1}
\ee
Here 
\be
\Delta\equiv\omega-\omega_0\,,\qquad\quad a\equiv (v/c)\omega_0\,.
\label{eq:notat1}
\ee
In the limit $v\to 0$ ($i.e.\ a\to 0$) the Lorentzian profile 
(\ref{eq:Lorentz}) is recovered. For small natural linewidths 
($\Gamma\ll |\Delta|,a$), one obtains  
\be
F(\Delta;a,\Gamma\ll |\Delta|)\simeq \frac{1}{2a}\left\{\begin{array}{l}
1, ~~|\Delta| \le a\\
0, ~~|\Delta|> a
\end{array}\right.\,, 
\label{eq:box}
\ee
that is, the line has a rectangular shape. Taking into account small 
natural widths of the initial and final atomic states would smooth  
the corners of the rectangle. 

Thus, the width of the spectral line is in this case fully 
determined by the Doppler broadening effect, despite the fact that the 
speed $v$ of the excited atom is fixed; the spread of the Doppler shifts 
is entirely due to the spread in the values of the angle $\theta$  
rather than due to the thermal motion of the excited ${\rm ^3He}$ atoms. 
The line profile is therefore different from the usual Voigt profile, 
which is a convolution of the Lorentzian and Gaussian distributions.

The parameter $a$ in eq.~(\ref{eq:notat1}) is the maximum Doppler shift 
of the photon frequency, corresponding to $\cos\theta=\pm 1$.  
It defines the positions of the edges of the rectangular line profile 
of the emitted photons. For a nonzero neutrino mass, this parameter 
decreases by $(\Delta v/c)\omega_0$ compared to the massless neutrino 
case. From eq.~(\ref{eq:Deltav}) it then follows that the relative change 
in the position of the spectral line's edge due to nonzero neutrino 
mass is  
\be
\frac{\Delta\omega_{\rm max}}{\omega_0}=\frac{\Delta v}{c}\simeq 
9.6\times 10^{-15}(m_\nu/{\rm eV})^2\,. 
\label{eq:edgeShift}
\ee

Because electron antineutrinos produced in BSBD are linear superpositions 
of neutrino mass eigenstates $\nu_k$, the lineshape of the emitted    
photon line is the weighted sum of the rectangles corresponding to 
different $\nu_k$, with weights given by the squared matrix elements of 
the leptonic mixing matrix, $|U_{ek}|^2$. The resulting photon 
line profile will therefore exhibit kinks (smoothed steps) at the 
frequencies corresponding to the onsets of the contributions of 
neutrinos of different masses. This is illustrated in Fig.~\ref{fig:box}, 
which shows the photon line profile for the case of one massless and one 
massive neutrino species in the limit of vanishingly small natural 
linewidths. This is a good first approximation for the case of a 
fourth, predominantly sterile, neutrino state, whose mass is 
significantly larger than those of the ordinary active neutrinos. 

Searching for a nonzero neutrino mass would then essentially amount to 
searching for kinks in the photon line profile.  
This is to some extent similar to looking for kinks in the electron 
spectra in direct neutrino mass measurement experiments based on CSBD, but 
instead of measuring electron energies in the keV range, our approach requires 
measuring photon frequencies corresponding  to eV-range energies. 
An important advantage of our method is that the fractional number of 
events $\Delta N_{\rm ph}/N_{\rm ph}$ in the $m_\nu$-sensitive region 
scales as $(a_0-a)/a_0\simeq m_\nu^2/2Q^2$. That is, 
\be
\frac{\Delta N_{\rm ph}}{N_{\rm ph}}\simeq \frac{m_\nu^2}{2Q^2}\simeq 
1.445\times 10^{-9}(m_\nu/{\rm eV})^2\,.
\label{eq:mnuSensit}
\ee
For comparison, in tritium CSBD experiments the fractional number of 
events in the mass-sensitive region scales as $(m_\nu/Q)^3$. 
Thus, the fractional number of useful events is increased by a very 
large factor of order $Q/m_\nu$ in our approach compared to the 
CSBD-based experiments. In particular, for $m_\nu=1$\,eV the fraction of 
useful photon events is $\sim 10^{-9}$ in the approach proposed here, 
whereas in BSBD of tritium the fraction of electrons with energies 
within the last ${\rm 1\,eV}$ below the endpoint energy is 
$\sim 10^{-13}$. 
 
\begin{figure}[!ht]
\begin{center}
\includegraphics[width=\linewidth]{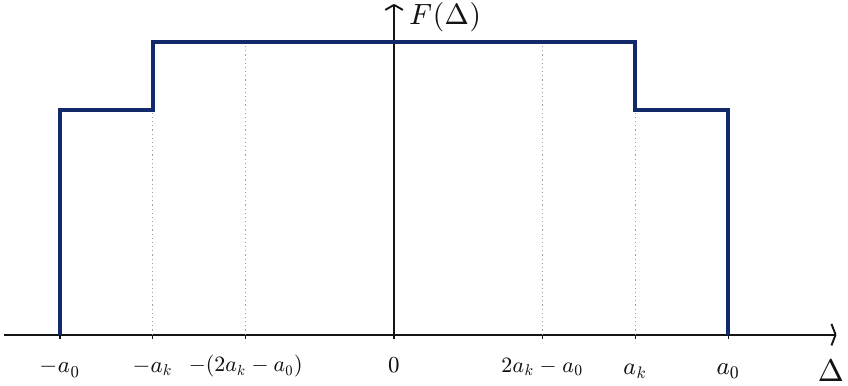}
\end{center}
\caption{\small Schematic representation of photon emission line 
spectrum in the case of one massless and one massive neutrino of mass 
$m_k$, in the limit of vanishingly small natural linewidth $\Gamma$ 
(not to scale). $\Delta=\omega-\omega_0$; 
the quantities $a_0$ and 
$a_k$ are the velocity parameters $a$ defined in (\ref{eq:notat1}) for 
$v=v_0$ and $v=v_k$, respectively. Grey dotted lines mark the borders 
of the central zone regions of the same widths as those of the wings. 
} 
\label{fig:box}
\end{figure}

\subsection{\label{sec:stat}Statistical 
requirements in the limit of vanishing natural linewidths}}

We shall now assess the statistical requirements for measuring 
neutrino masses within our approach. We do this first in a 
simplified and idealized manner, assuming that the effects of thermal 
motion of the parent tritium atoms and the natural linewidths of the 
relevant atomic states of ${\rm ^3He}$ can be neglected. 

Consider a ${\rm ^3He}$ emission line in the case of two neutrino 
species: one massless neutrino and one massive neutrino of mass $m_k$ 
(see Fig.~\ref{fig:box}). 
We denote by $a_0$ and $a_k$ the velocity parameters $a$ defined in 
(\ref{eq:notat1}) for $v=v_0$ and $v=v_k$, respectively. In the wings  
of the photon line profile, i.e.\ in the regions $a_k<|\Delta|<a_0$, only 
massless neutrinos contribute to the emitted photon spectrum.
In the central region, corresponding to $|\Delta|<a_k$, both massless 
and massive neutrinos contribute, with the weights $1-|U_{ek}|^2$ and 
$|U_{ek}|^2$, respectively. 

Let $N$ be the full number of the observed photon events corresponding 
to the atomic transition under discussion. 
The numbers of events corresponding to the wings and to the central region of the line 
profile, which we denote by $N_w$ and $N_c$ respectively, are 
proportional to the integrals of the photon spectrum over these regions, 
i.e.\ to the areas of the corresponding parts of the line profile 
shown in Fig.~\ref{fig:box}. 

To estimate the statistical significance of the neutrino mass determination, 
we consider the number $N_c'$ of observed photons in the central region of the line 
profile, but restricted to 
the intervals in $\Delta$ of the same widths as 
the wings, i.e.\ in the ranges $2a_k-a_0<|\Delta|<a_k$. 
In the limit $m_k=0$, the spectrum of the studied line is purely 
rectangular, so the numbers of events in the spectral intervals 
of equal widths must coincide.  
A nonzero difference $N_c'-N_w \,\simeq\,
N\frac{\Delta v}{v_0}|U_{ek}|^2$ would therefore signify nonvanishing 
neutrino mass. The statistical significance of observing $m_k\ne 0$ 
(the number of standard deviations) is then 
\be
Z~=~\frac{N_c'-N_w}{\sqrt{N_c'+N_w}}~\simeq
~\frac{\sqrt{N\frac{\Delta v}{v_0}}
|U_{ek}|^2}{\sqrt{2-|U_{ek}|^2}}.
\label{eq:Z}
\ee
(see Appendix~\ref{sec:appStat} for details). 
The number of observed events required for a significance $k\sigma$ 
is thus 
\be
N~\simeq~k^2\,\frac{2-|U_{ek}|^2}{(\Delta v/v_0)|U_{ek}|^4}~=~
k^2\frac{4Q^2}{m_k^2 |U_{ek}|^4}\big(1-\frac{1}{2}|U_{ek}|^2\big).
\label{eq:N}
\ee

Consider a sterile neutrino with mass 1\,keV and the mixing 
parameter $|U_{ek}|^2$ equal to 
$10^{-3}$,\,$10^{-4}$ or $10^{-6}$. To achieve a 3$\sigma$ 
significance in the observation of a nonzero neutrino mass, the required 
number of observed photons is then, respectively, 
\be 
N\simeq 1.2\times 10^{10}\,;\qquad 1.2\times 10^{12}\,;\qquad 
1.2\times 10^{16}\,. 
\label{eq:Nreq1} 
\ee 
As we have ignored here the effects of finite natural linewidth and 
any possible systematic errors, the numbers in eq.~(\ref{eq:Nreq1})  
should be considered as a minimum requirement.
 
Let the overall photon detection efficiency for the line under study 
be $1/\eta$, which means that only a fraction $1/\eta$ of the 
emitted photons is actually detected. For tritium BSBD transitions to 
the $1s3s$ state of triplet helium and a tritium source with an activity 
of $10^{12}\,{\rm Bq}$ (2.8\,mg of tritium), the times required to 
accumulate the numbers of events given in~(\ref{eq:Nreq1}) are, 
respectively, 
\be
{\rm 0.9\eta\;min\,;\qquad\quad ~~1.5\eta\,\; hours\,;
\qquad
1.7\eta\; yr}.
\label{eq:treq1}
\ee
If the sterile neutrino mass is 
200\,eV rather than 1\,keV, the required event numbers and 
the running times of the experiment must be multiplied by a factor 
of 25. 

The above approach can be readily extended to the case of 
more than two neutrino species. Consider the ordinary active neutrinos 
with mass eigenvalues $m_1$, $m_2$ and $m_3$. The relevant mixing 
parameters in this case are  
\be
|U_{e1}|^2\simeq 0.7,\qquad |U_{e2}|^2\simeq 0.3,\qquad
|U_{e3}|^2\simeq 0.022\,.
\label{eq:actMix}
\ee
In Table~\ref{tab:masses} the values of the masses of active neutrinos 
used in our estimates are shown for the normal (NO) and inverted (IO) 
mass orderings, assuming hierarchical mass spectrum. 
In this case the masses of the lightest neutrinos are too small to be 
probed by direct mass measurement experiments. Also, for IO, the 
masses of the heaviest and next-to-heaviest active neutrinos are 
too close to each other to be resolved. 
 
\begin{table*}[h]
\centering
\begin{tabular}{|c|c|c|c|}
\hline
\hline
 & $m_1 \,({\rm eV})$ & $m_2\,({\rm eV})$ &
 $m_3\,({\rm eV})$ \\
 \hline
 NO & $\simeq 0$ & $8.7\times 10^{-3}$ & $ 0.05$ \\
\hline
IO &
$0.0492$ & $0.05$ & $\simeq 0$\\
\hline
\end{tabular}
\caption{{\small Values of masses of active neutrinos used in our 
sensitivity estimates, assuming hierarchical mass spectrum. NO and IO 
stand for the normal and inverted mass orderings, respectively.}}
\label{tab:masses}
\end{table*}
Using eqs.~(\ref{eq:N}), (\ref{eq:actMix}) together with the 
neutrino masses given in Table~\ref{tab:masses}, one can determine the 
numbers of observed events and the corresponding running times of the 
experiment required to observe a nonzero neutrino mass 
at the $3\sigma$ significance level. For NO, one obtains
\bea
&N(m_2)\simeq 1.8\times 10^{15}\,,\qquad & 
t(m_2)\simeq 0.25\eta\,{\rm yr} \\
&N(m_3)\simeq 1.0\times 10^{16}\,,\qquad &t(m_3)\simeq 1.4\eta\,
{\rm yr}\,,
\eea
whereas for IO, 
\be
N(m_{\nu_1+\nu_2})\simeq 5.4\times 10^{13}\,,\qquad
t(m_{\nu_1+\nu_2})\simeq 64\eta \,{\rm hours}
\vspace*{2mm}
\ee
As before, the values of the running time 
correspond to BSBD transitions to the triplet $1s3s$ excited state of 
helium atoms and a tritium source with an activity of $10^{12}$\,Bq.  

The simplified consideration presented in this subsection applies to the 
statistical significance of establishing the existence of kinks in the 
profile of the studied line, signifying nonzero neutrino masses. 
The accuracy of measurements of the positions of these kinks, which 
is essentially the accuracy of determination of the underlying neutrino 
masses, is determined by the frequency resolution of the lineshape 
measurements [see eq.~(\ref{eq:edgeShift})].

\subsection{\label{sec:width}\!Effects of finite natural widths of the 
excited states of ${\rm ^3He}$}

The profiles of the photon lines emitted by ${\rm ^3He}$ atoms take an 
approximately rectangular shape (in the rest frame of the parent tritium 
atoms) when the sum $\Gamma$ of the natural linewidths of the initial 
and final atomic states of helium is small compared to the maximum 
Doppler shift $a=(v/c)\omega_0$. However, when analyzing the line 
profiles to extract the neutrino masses, the condition for 
neglecting the natural linewidths is much more 
stringent: $\Gamma$ must be small not only compared to $a$, but actually 
compared to the deviation of $a$ from its value in the limit of massless 
neutrinos: 
\be
\Gamma \ll a_0-a=\frac{\Delta v}{v_0}a_0\,.
\label{eq:condit1}
\ee
What happens if this condition is not met? As a concrete example, 
consider the red-line transition \,$1s3s\to 1s2p$ \,in triplet helium. 
For this transition, 
\be
\omega_0=1.755\,{\rm eV}\,;\quad
\Gamma_i=1.83\times 10^{-8}\,{\rm eV}\,;\quad
\Gamma_f=6.72\times 10^{-9}\,{\rm eV}\,;\quad
\Gamma=\Gamma_i+\Gamma_f=2.5\times 10^{-8}
\,{\rm eV}\,.
\label{eq:transit1a}
\ee
The maximum Doppler shifts for a massless neutrino and for a neutrino 
with mass $m_k=1$\,keV are 
\be
a_0=\frac{v_0}{c}\omega_0=1.16\times 10^{-5}\,{\rm eV};~~\quad
a_k=
a_0(1-1.45\times 10^{-3})\,.
\label{eq:transit1b}
\ee
Thus, although $\Gamma$ is much smaller than $a_0$, it is comparable 
to the difference $a_0-a_k\simeq 1.7\times {\rm 10^{-8}\;eV}$. 
Figure~\ref{fig:widths} shows the photon spectrum obtained from
eqs.~(\ref{eq:Dopp1}) and (\ref{eq:notat1}) using the physical natural
linewidth $\Gamma=2.5\times10^{-8}\,$eV in both panels.  The left panel
shows the full spectrum over both Doppler edges, whereas the right panel
shows the region near the high-frequency edge,
$\Delta\in[a_0-3\Gamma,\,a_0+3\Gamma]$. 
The red and blue curves correspond to a massless neutrino and a neutrino 
with mass 1\,keV, respectively. One can see that the difference between 
the massless and massive neutrino cases is still quite noticeable, 
though distinguishing between them may require rather high event 
statistics. 

The approximations entering the calculation are those stated above: 
the parent tritium atom is taken to be at rest, Doppler shifts are 
retained to first order in $v/c$, recoil directions are 
isotropically averaged, and thermal broadening, collisional 
de-excitation, detector resolution, and instrumental frequency 
calibration errors are not included.  These effects must be added 
separately before the profile can be interpreted as a complete 
experimental response function.

\begin{figure}[!ht]
\centering
\includegraphics[width=\textwidth]{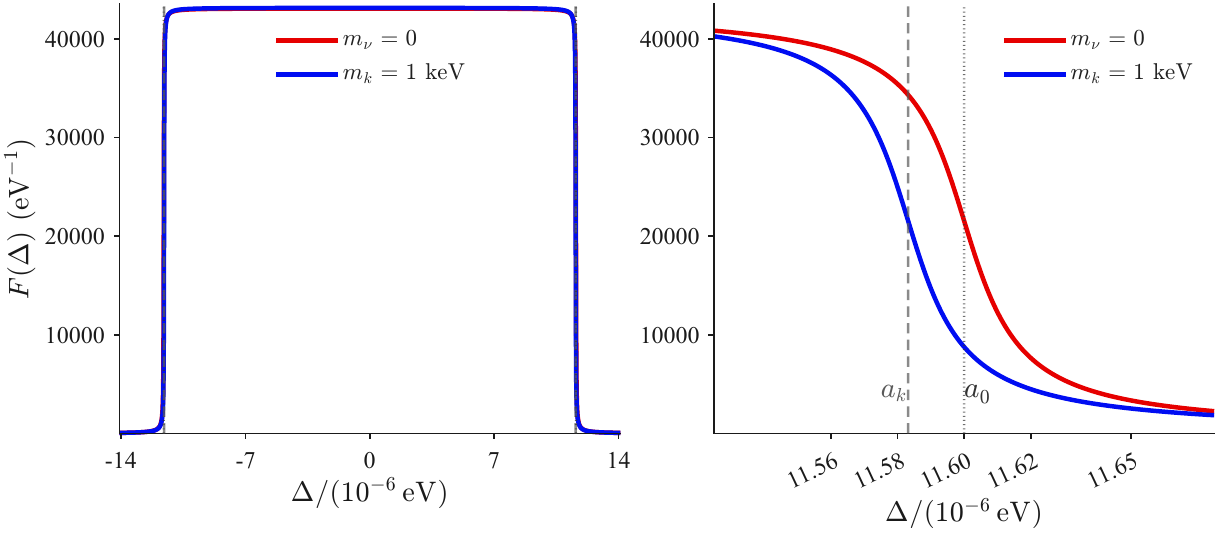}
\caption{
\small Frequency spectrum of photons from the $1s3s\to 1s2p$ 
transition in triplet helium. 
Red and blue curves correspond to a massless neutrino and a 
neutrino mass eigenstate with  $m_k=1$\,keV, respectively. Left panel: 
full spectrum over both Doppler edges, 
$\Delta\in[-a_0-100\Gamma,\,a_0+100\Gamma]$; right panel: zoom on the
high-frequency endpoint, 
$\Delta\in[a_0-3\Gamma,\,a_0+3\Gamma]$. Both panels use
$\Gamma=2.5\times 10^{-8}$\,eV. Including theoretical or experimental
natural-width uncertainties (0.01\% or 3\%, respectively) leads   
to the uncertainty bands that are narrower than, or comparable to, the 
plotted line thickness on this scale.
}
\label{fig:widths}
\end{figure}

\subsection{\label{sec:statFinite}Statistical requirements with finite 
linewidths}

The box-counting estimate introduced in Sec.~\ref{sec:stat} provides a 
useful starting point because it directly shows how the statistical 
sensitivity depends on $|U_{ek}|^2$ and $m_k^2$.  However, it applies 
when the linewidth $\Gamma$ is smaller than, or comparable to, the 
mass-induced edge displacement $a_0-a_k$.  For light-neutrino mass 
measurements, this condition is not necessarily satisfied, and a more 
refined statistical treatment is therefore required. For a finite 
natural linewidth, the relevant normalized profile is the function 
$F(\Delta;a,\Gamma)$ in eq.~(\ref{eq:Dopp1}). We define 
the mixed two-component spectrum as
\be
F_{\rm mix}(\Delta)=
\left(1-|U_{ek}|^2\right)F(\Delta;a_0,\Gamma)
 +|U_{ek}|^2 F(\Delta;a_k,\Gamma)\,,
\label{eq:finiteStatMix}
\ee
and compare it with the zero-neutrino-mass null spectrum 
\be
F_0(\Delta)=F(\Delta;a_0,\Gamma)\,,
\label{eq:finiteStatNull}
\ee
where $|U_{ek}|^2$ is the electron-flavor admixture of the massive 
eigenstate.  The observable deformation is therefore the ratio
\be
\frac{F_{\rm mix}(\Delta)}{F_0(\Delta)}=
\left(1-|U_{ek}|^2\right)+|U_{ek}|^2
\,\frac{F(\Delta;a_k,\Gamma)}
              {F(\Delta;a_0,\Gamma)}\,.
\label{eq:finiteStatRatio}
\ee
This ratio, rather than the absolute profile alone, most directly 
reveals the spectral distortion induced by nonzero neutrino masses 
relative to the reference spectrum obtained by setting all neutrino 
masses to zero.

We use a Fisher-integral estimator to convert a fixed spectral 
deformation into the number of detected photons required for a target 
statistical significance.  Let $F_0(\Delta)$ be the normalized null 
spectrum and $F_1(\Delta)$ the normalized alternative 
spectrum that takes neutrino masses into account, with 
\(\int F_i(\Delta)\,d\Delta=1\).  If \(N\) photons are detected in the 
line, the expected counts in an interval \(d\Delta\) are 
\(N F_i(\Delta)d\Delta\).  For Poisson statistics and 
small bin-by-bin differences, the leading likelihood expansion gives
\be
Z_{\rm F}^2 \simeq N\,{\cal I}\,,\qquad
{\cal I}=
\int d\Delta\,
\frac{\left[F_1(\Delta)-F_0(\Delta)\right]^2}
     {F_0(\Delta)}\,,
\label{eq:finiteStatFisher}
\ee
where \(Z_{\rm F}\) is the expected Gaussian-equivalent statistical 
separation for the fixed pair of spectra.  The required statistics are then $
N_{\rm req}(Z_{\rm F})={Z_{\rm F}^2}/{{\cal I}}\,.
\label{eq:finiteStatNreq}
$
The integral ${\cal I}$ is dimensionless; all finite-linewidth 
applications of the estimator below include both Doppler edges unless 
stated otherwise.

\begin{figure}[!ht]
\centering
\includegraphics[width=0.96\textwidth]{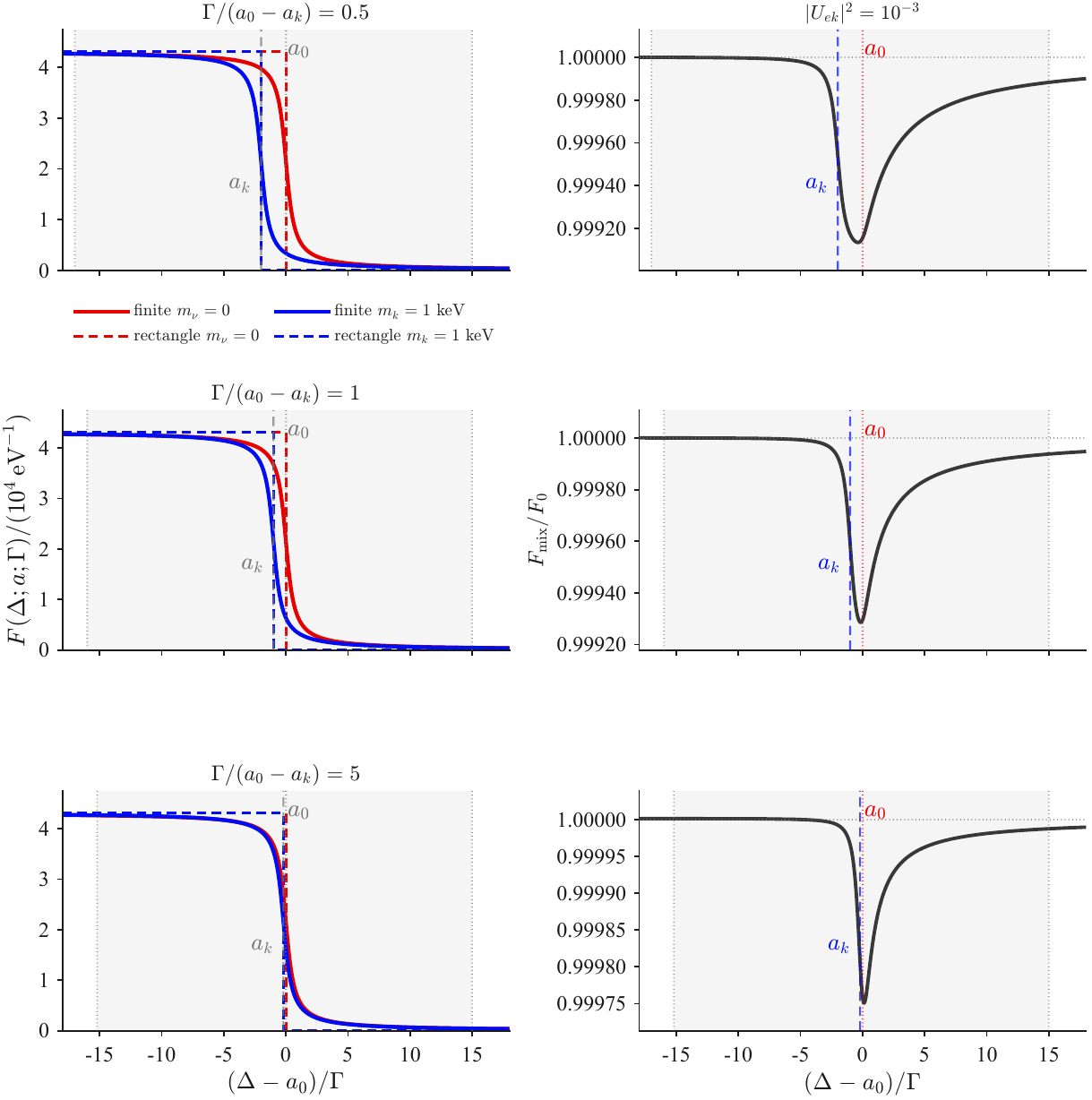}
\caption{\small Finite-linewidth spectra entering the Fisher-integral 
estimator for a benchmark $m_k=1$\,keV neutrino.  Rows correspond to 
$\Gamma/(a_0-a_k)=0.5,1,5$.  The left panels show the finite-linewidth
spectra and ideal rectangle limits; the right panels show the observable ratio 
$F_{\rm mix}/F_0$ for $|U_{ek}|^2=10^{-3}$; the shaded 
region is the one-edge integration interval.  The quoted statistics use 
both Doppler edges by symmetry.}
\label{fig:finite_linewidth_fisher_profiles}
\end{figure}

For $m_k=1$\,keV, $|U_{ek}|^2=10^{-3}$, and $Z_{\rm F}=3$ (which corresponds to 3$\sigma$ significance), the 
box-counting estimate of Sec.~\ref{sec:stat}  gives 
$N_{\rm req}=1.24\times10^{10}$ detected photons.  Evaluating 
Eq.~(\ref{eq:finiteStatFisher}) for the corresponding zero-linewidth 
rectangular spectra over the full profile gives 
$N_{\rm req}=6.20\times10^9$.  Replacing the rectangular spectra by the 
finite-linewidth profiles \(F(\Delta;a,\Gamma)\) of 
Eq.~(\ref{eq:Dopp1}) in Eq.~(\ref{eq:finiteStatMix}) 
and including both Doppler edges gives 
$N_{\rm req}=9.89\times10^9$, $1.35\times10^{10}$, and 
$4.54\times10^{10}$ for 
$\Gamma/(a_0-a_k)=0.5$, $1$, and $5$, respectively. 
Thus, the box-counting estimate remains close to the Fisher-integral estimator as long as the mass-induced spectral features remain resolved or only weakly broadened, but deviates once $\Gamma$ becomes significantly larger than $a_0-a_k$.
For the triplet $1s3s\to1s2p$ transition discussed in the previous 
subsection, $a_0-a_k=1.68\times10^{-8}$\,eV for $m_k=1$\,keV, while 
$\Gamma=2.5\times10^{-8}$\,eV.  Thus 
$\Gamma/(a_0-a_k)=1.49$, so the edge is moderately smeared but not lost.  
For $|U_{ek}|^2=10^{-3}$ and $Z_{\rm F}=3$, the box-counting estimate 
gives $N_{\rm req}=1.24\times10^{10}$ detected photons, whereas the 
application of the Fisher-integral estimator to both Doppler edges gives 
$N_{\rm req}=1.72\times10^{10}$.  The box approximation provides a 
reasonable statistical scale for this transition, while the 
Fisher-integral estimator requires a slightly larger exposure because 
it accounts for finite-linewidth smearing through 
Eq.~(\ref{eq:finiteStatRatio}).

For active-neutrino masses, the edge displacements are much smaller 
than the triplet $1s3s\to1s2p$ natural linewidth, so the statistical 
requirement cannot be assessed from resolved intervals.  The relevant 
observable is instead the finite-linewidth spectral ratio, and the 
appropriate sensitivity estimator is the Fisher integral of 
Eq.~(\ref{eq:finiteStatFisher}).  For three active neutrinos, the 
two-component spectrum in Eq.~(\ref{eq:finiteStatMix}) is replaced by 
an incoherent sum over the three mass eigenstates, with coefficients 
$|U_{ek}|^{2}$ fixed by oscillation data. 
Figure~\ref{fig:three_active_io_10mev_fisher} compares the resulting 
spectrum with the reference spectrum obtained by setting all three 
neutrino masses to zero. For illustration, we consider inverted 
ordering with $m_{\rm lightest}=m_3=10.0,\mathrm{meV}$, $\Delta 
m_{21}^{2}=7.41\times10^{-5},\mathrm{eV}^{2}$, and $|\Delta 
m_{31}^{2}|=2.44\times10^{-3},\mathrm{eV}^{2}$. This gives 
$(m_1,m_2,m_3)=(50.4,51.1,10.0),\mathrm{meV}$ and corresponding edge 
shifts of $(43,44,1.7)\times10^{-18},\mathrm{eV}$. Because these shifts 
are approximately nine orders of magnitude smaller than the natural 
linewidth of the triplet transition, the ideal box approximation is 
inadequate and the Fisher-integral estimator must be used. Integrating 
the Fisher density over finite windows around both Doppler edges yields 
a statistical requirement of $N^{\Gamma}(3\sigma)=1.9\times10^{21}$ 
detected photons.

\begin{figure}[!ht]
\centering
\includegraphics[width=0.92\textwidth]{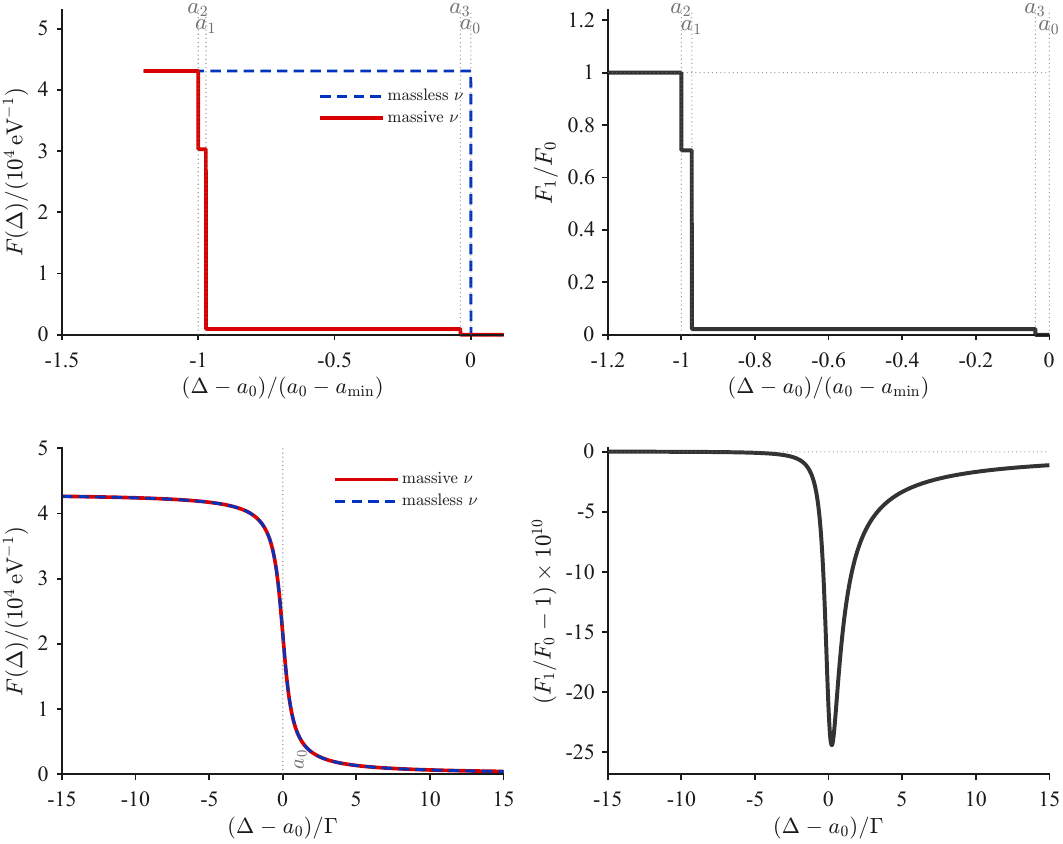}
\caption{\small Comparison of the predicted photon spectra for three active neutrinos
with inverted mass ordering and $m_3 = 10\,\mathrm{meV}$, relative to
the reference case of three massless neutrinos. For the inverted ordering 
considered here, the lowest kinematic edge is $a_{\min}=a_2$. 
The upper-left panel shows the ideal rectangular spectra, and the upper-right 
panel shows their ratio. The lower-left panel shows the corresponding 
finite-linewidth spectra $F(\Delta)$ for the triplet-helium $1s3s\to1s2p$ 
transition, while the lower-right panel shows the relative deviation of their 
ratio from unity. Using the Fisher-integral estimator, the photon 
count required for a $3\sigma$ statistical discrimination is 
$1.9\times10^{21}$.}
\label{fig:three_active_io_10mev_fisher}
\end{figure}

All event counts quoted in this subsection represent purely statistical lower 
bounds. A complete experimental sensitivity assessment must also account for 
thermal broadening, detector response, frequency calibration, acceptance, 
backgrounds, and correlations among nuisance parameters. The statistical 
requirements therefore appear feasible for keV-scale sterile-neutrino 
benchmarks when the spectral edges remain resolved or only weakly broadened. 
By contrast, in the active three-neutrino case for the physical triplet 
red-line channel, the natural linewidth strongly suppresses the observable 
spectral deformation; control of systematic uncertainties will therefore be 
crucial for any realistic sensitivity assessment.

\subsection{\label{sec:thermal}\!\!Effects of thermal motion of parent
tritium atoms}

The one-to-one correspondence between the neutrino mass and the velocity of
the produced ${\rm ^3He}$ atoms holds only in the rest frame of the parent
tritium atoms.  In the laboratory frame, thermal motion of tritium must be
taken into account.  We therefore consider how this motion modifies the
photon line profile.  Let $\vec v=v_k\hat v$ be the helium recoil
velocity in the rest frame of the parent tritium atom, $\vec v_T$ the parent
velocity in the laboratory, and $\hat{\mathbf n}_\gamma$ the detected photon
direction.
In the laboratory, the helium atom inherits the translational velocity of
the parent atom in addition to its BSBD recoil velocity.  To first order in
the velocities, the Doppler detuning of a photon is proportional to the
component of the total helium velocity along the detected photon direction.
Since this laboratory velocity is $v_k\hat v+\vec v_T$, the leading Doppler
detuning is
\be
\Delta\simeq\frac{\omega_0}{c}
\left(v_k\cos\theta+v_{T\parallel}\right),
\qquad
\cos\theta\equiv\hat v\cdot\hat{\mathbf n}_\gamma,
\qquad
v_{T\parallel}\equiv\vec v_T\cdot\hat{\mathbf n}_\gamma .
\label{eq:thermalDetuningNew}
\ee
For an isotropic Maxwell--Boltzmann distribution, $v_{T\parallel}$ is a
Gaussian random variable with
\be
\langle v_{T\parallel}\rangle=0,\qquad
\langle v_{T\parallel}^2\rangle
=\frac{k_BT}{M_i}=\frac{v_{th{\rm T}}^2}{3}.
\label{eq:thermalProjectionNew}
\ee
Here $v_{th{\rm T}}=\sqrt{3k_BT/M_i}$ is the three-dimensional root-mean-square
(rms) thermal speed.  The leading thermal photon frequency width is therefore
\be
\sigma_{\rm th}=\frac{\omega_0}{c}
\sqrt{\frac{k_BT}{M_i}}
=\frac{\omega_0}{c}\frac{v_{th{\rm T}}}{\sqrt{3}}\,.
\label{eq:thermalWidthNew}
\ee
We first neglect the natural linewidth, $\Gamma=0$, in order to isolate the
effect of finite temperature.  If $G_T(\epsilon)$ denotes the normalized
Gaussian with rms width $\sigma_{\rm th}$, the rectangular profile in
Eq.~(\ref{eq:box}) is replaced by
\be
P_{T,k}^{(0)}(\Delta)=\int_{-\infty}^{\infty}d\epsilon\,
F_0(\Delta-\epsilon;a_k)G_T(\epsilon)
=\frac{1}{4a_k}\left[
{\rm erf}\!\left(\frac{\Delta+a_k}{\sqrt{2}\sigma_{\rm th}}\right)
-{\rm erf}\!\left(\frac{\Delta-a_k}{\sqrt{2}\sigma_{\rm th}}\right)
\right] .
\label{eq:thermalGaussianNew}
\ee
Here $F_0(\Delta;a_k)$ denotes the zero-linewidth rectangular spectrum.
Thus, finite temperature smooths both kinematic edges with a width that
scales as $\sqrt{T}$, but it does not shift the line center or the kinematic
edge reference positions $\Delta=\pm a_k$.

A useful dimensionless measure of the thermal smearing is
${\cal R}_T\equiv(a_0-a_k)/\sigma_{\rm th}$, where $a_0-a_k$ is the inward
displacement of one edge caused by the
neutrino mass, while $\sigma_{\rm th}$ is the one-standard-deviation thermal
width.  Values ${\cal R}_T\ll1$ indicate that the mass-dependent edge
displacement is small compared with the thermal smoothing scale.

Figure~\ref{fig:thermalEdgeNew} illustrates this effect for atomic tritium at
$T=1\,$K, comparing $m_k=0$ and $m_k=1\,$keV.  At this temperature,
$\sigma_{\rm th}=0.31\,\mu{\rm eV}$, whereas the mass-induced displacement
of one edge is only $a_0-a_k=0.017\,\mu{\rm eV}$, giving
${\cal R}_T=0.055$.  Thermal motion therefore
strongly smooths the mass-dependent structure, in direct analogy with the
degradation caused by a finite natural linewidth discussed in
Sec.~\ref{sec:width}.  If the temperature is known exactly, the broadened
hypotheses remain distinct, and a larger photon sample can in principle reach
a chosen statistical significance, although the required event count may
become prohibitively large.  Uncertainties in the temperature or detector
response can introduce additional degeneracies that cannot in general be
overcome by statistics alone.

\begin{figure}[!th]
\centering
\includegraphics[width=0.75\textwidth]{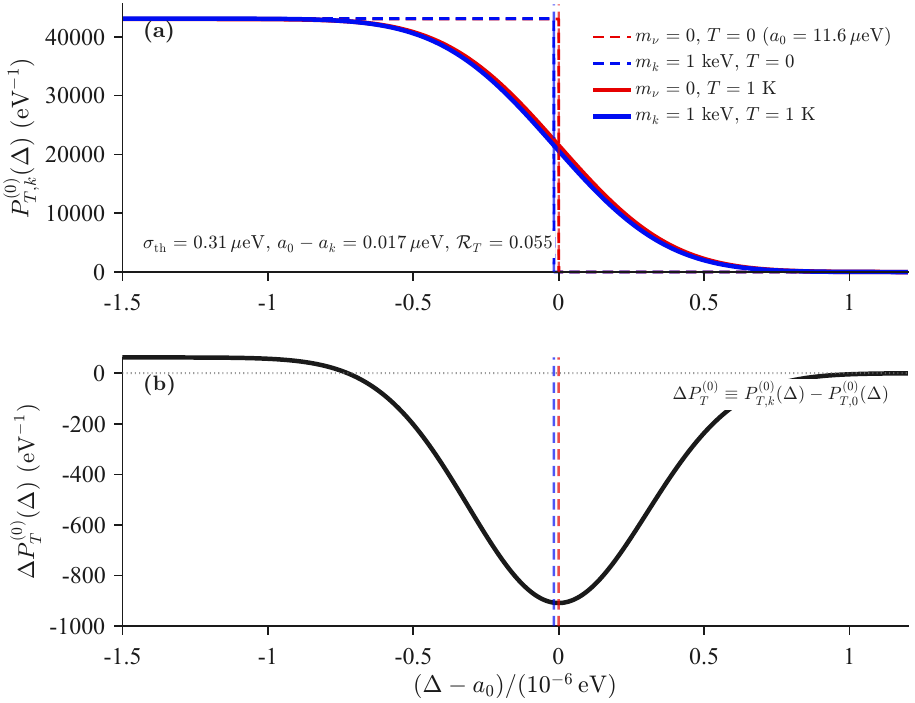}
\caption{\small High-frequency edge of the normalized
atomic-tritium BSBD photon spectrum for $m_k=0$ and
$m_k=1\,$keV. Dashed curves and vertical lines show the ideal $T=0$,
$\Gamma=0$ rectangular-spectrum edges; solid curves show the leading
Gaussian thermal broadening at $T=1\,$K, also with $\Gamma=0$, in order to
isolate the temperature effect. The lower panel gives the absolute
difference between the two thermally broadened spectra. For this benchmark,
$\sigma_{\rm th}=0.31\,\mu{\rm eV}$,
$a_0-a_k=0.017\,\mu{\rm eV}$, and
${\cal R}_T\equiv(a_0-a_k)/\sigma_{\rm th}=0.055$; thus ${\cal R}_T$
is the ratio of the mass-induced displacement of one edge to the thermal
rms width.}
\label{fig:thermalEdgeNew}
\end{figure}

We now include the natural linewidth.  The finite radiative lifetimes of the
initial and final helium states give the emitted photon a Lorentzian frequency
distribution, whereas the line-of-sight thermal velocity of the parent atom
produces the Gaussian Doppler shift described above.  These two types of
fluctuations have distinct physical origins and are statistically independent
in the present approximation.  Their probability distributions must therefore
be convolved, so that the finite-temperature profile is
\be
P_{T,k}(\Delta)=\int_{-\infty}^{\infty}d\epsilon\,
F(\Delta-\epsilon;a_k,\Gamma)G_T(\epsilon)\,.
\label{eq:thermalConvolutionNew}
\ee
The complete profile therefore results from applying both the natural
Lorentzian broadening and the thermal Gaussian broadening to the rectangular
recoil spectrum.  It is not a single Voigt peak, although each kinematic edge
is broadened with a Voigt shape.
Because both broadening distributions are centered and symmetric, they smooth
the kinematic edges without shifting their reference positions
$\Delta=\pm a_k$; neither effect therefore produces an intrinsic displacement
of the edge location.

For the $m_k=1\,$keV benchmark, a useful temperature scale can be defined by
requiring the thermal Gaussian FWHM not to exceed the natural Lorentzian
FWHM.  For a Gaussian proportional to
$\exp[-\epsilon^2/(2\sigma_{\rm th}^2)]$, the half-maximum occurs at
$|\epsilon|=\sqrt{2\ln2}\,\sigma_{\rm th}$.  Its full width at half maximum
is therefore $\Gamma_{\rm G}^{\rm th}=2\sqrt{2\ln2}\,\sigma_{\rm th}$.
The natural linewidth $\Gamma$ used in Eq.~(\ref{eq:Lorentz}) is also a full
width at half maximum.  A like-for-like comparison therefore requires
$\Gamma_{\rm G}^{\rm th}\leq\Gamma$.
For the red-line transition of Eq.~(\ref{eq:transit1a}), with
$\omega_0=1.755\,$eV and $\Gamma=2.5\times10^{-8}\,$eV, this condition
gives
\be
T\leq T_{\rm max}=\frac{M_i}{k_B}
\left[\frac{\Gamma}
{2\sqrt{2\ln2}\,\omega_0}\right]^2
=1.2\,{\rm mK}\,.
\label{eq:thermalTmaxNew}
\ee
This value is independent of the neutrino mass; the choice $m_k=1\,$keV only
specifies the kinematic edge illustrated below.  The criterion prevents the
thermal Gaussian FWHM from exceeding the intrinsic natural width, but it does
not make the thermal contribution negligible.  As shown in
Fig.~\ref{fig:thermalNaturalNew}, at $T=T_{\rm max}$ the combined box--Voigt
edge is broader than either the natural-only or thermal-only edge.

\begin{figure}[!ht]
\centering
\includegraphics[width=0.8\textwidth]{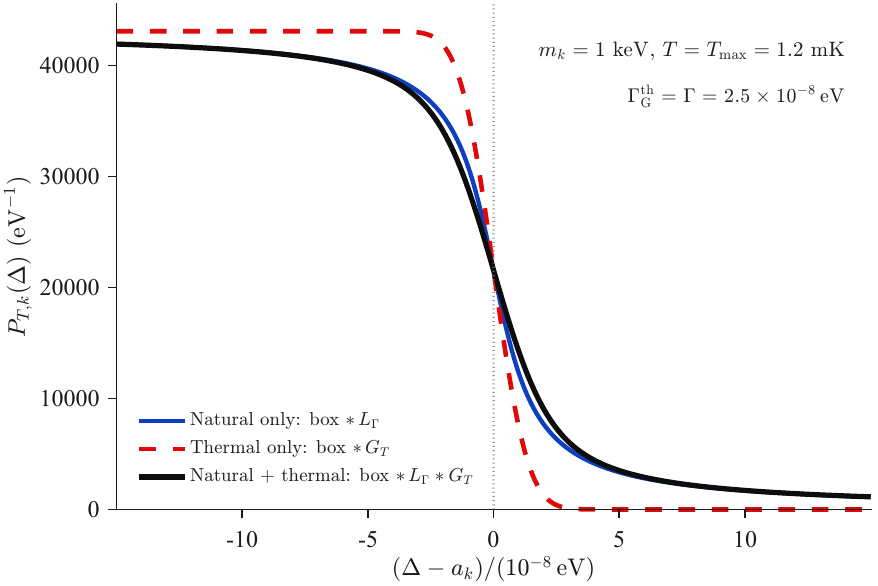}
\caption{\small High-frequency edge for $m_k=1\,$keV at
$T=T_{\rm max}=1.2\,$mK. The blue curve includes only the natural
Lorentzian width, the red dashed curve includes only Gaussian thermal
broadening, and the black curve includes both effects through
Eq.~(\ref{eq:thermalConvolutionNew}). At this temperature,
$\Gamma_{\rm G}^{\rm th}=\Gamma=2.5\times10^{-8}\,$eV. Equality of the two
individual FWHM values does not imply that the combined profile has the same
width as either component.}
\label{fig:thermalNaturalNew}
\end{figure}

To quantify the corresponding statistical degradation in the same idealized
framework, we apply the Fisher-integral estimator of
Eq.~(\ref{eq:finiteStatFisher}) to the three-active-neutrino inverted-ordering
benchmark considered in Sec.~\ref{sec:statFinite}, now including both the
natural linewidth and thermal Doppler broadening at the fixed temperature
$T=1.2\,$mK.  The resulting kinematic edges have a Voigt shape.  Integrating
the Fisher density over the same finite windows around both Doppler edges
gives $N^{\Gamma+T}(3\sigma)=2.3\times10^{21}$ detected photons, compared with
$N^{\Gamma}(3\sigma)=1.9\times10^{21}$ for the natural-linewidth-only case.
Thus, for this benchmark and an exactly known temperature, thermal broadening
increases the purely statistical event requirement by approximately $21\%$.
At $90\%$ C.L. ($Z_{\rm F}=1.64$), the corresponding requirement is
$6.9\times10^{20}$ detected photons.  Using the red-photon production rate of
Eq.~(\ref{eq:Gamma2}) and three years of continuous operation, this requires a
constant source inventory of $1.8\times10^{25}$ tritium atoms, equivalent to
approximately $88\,$g of atomic tritium and $2.9\times10^{5}$ times the
inventory of the KATRIN WGTS source\footnote{\normalfont The current KATRIN
WGTS source, containing approximately $6\times10^{19}$ tritium atoms, would
provide an ${\cal O}(1\,{\rm eV})$ sensitivity within this idealized
statistical framework.}.

This linewidth criterion is only a useful scale benchmark, not an
experimental sensitivity calculation.  The neutrino mass
must be inferred by fitting the position and shape of the measured edge.  A
quantitative sensitivity and event requirement can only be obtained after
specifying the detector response and the relevant systematic effects.  They
depend on the photon event statistics, the natural linewidth, the
instrumental frequency resolution, frequency calibration, backgrounds,
acceptance, the actual experimental temperature, and the precision with
which this temperature is independently known.  Such a setup-dependent
joint line-shape analysis is beyond the scope of this work.

\section{
\label{sec:summary}
Summary and discussion}
\label{sec:disc}

We have proposed a method for the experimental observation of the BSBD 
of tritium based on the detection of photons emitted by the neutral 
${\rm ^3He}$ atoms produced in this process. The key point is that it  
populates a few low-lying excited states of ${\rm ^3He}$ with significant 
probabilities. The subsequent radiative decay of these states produces 
photons that can serve as a clear signature of BSBD. 

We have identified the most promising atomic transitions and transition 
sequences for this purpose: these are the red 
lines in the spectra of singlet and triplet helium, the NIR line in 
triplet helium, and their combinations. We have also analyzed possible 
backgrounds to this detection method and ways to suppress them.

We have proposed a novel method for direct neutrino mass measurement 
based on the detection of atomic photons emitted by ${\rm ^3He}$ atoms 
produced in BSBD of tritium. The two-body kinematics of the final state 
of this process means that, for a given neutrino mass, the emitted 
neutrino and the helium atom are monoenergetic. The velocity of the 
helium atom depends weakly on the neutrino mass, so its accurate 
determination would constitute a neutrino mass measurement. 

We propose to measure the velocities of the ${\rm ^3He}$ atoms by 
observing the 
Doppler broadening of their photon emission lines. 
Since electron antineutrinos are linear superpositions of neutrino mass 
eigenstates with different masses,
the spectra of the emitted ${\rm ^3He}$
photons are composite; they must exhibit characteristic kinks 
corresponding to the onsets of contributions of different neutrino masses. 
A measurement of the neutrino mass would therefore 
require an accurate 
measurement of the 
photon line profiles. Thus, our approach 
is based on atomic spectroscopy rather than on measurements of 
electron energy spectra, as employed by the conventional CSBD-based direct 
neutrino mass measurement experiments. 

As can be seen from eq.~(\ref{eq:edgeShift}), the currently achieved 
accuracy in photon line profile measurements, ranging from $\sim 10^{-4}$ 
to $10^{-11}$, implies that neutrino masses in the range from 
$\sim$18.6\,keV down to 30\,eV can be probed. Note, however, that the 
progress in atomic spectroscopy is very rapid, and significant 
improvements can be expected in the near future.  

It should be noted that we used the linear (nonrelativistic) Doppler effect 
approximation to analyze the sensitivity of our approach to neutrino masses. This 
approximation remains valid as long as the relativistic corrections to the Doppler 
shifts are smaller than the distortions of the photon 
lineshapes induced by a nonzero neutrino mass. This condition holds  
for $m_\nu\gtrsim (Q^3/M)^{1/2}\simeq 50$\,eV. To probe smaller neutrino 
masses, the full relativistic expression for the Doppler frequency shifts 
must be used. This is completely straightforward and does not lead 
to any loss of neutrino mass sensitivity of our method.  

We have studied the effect of finite natural linewidths on the photon line
profiles.  Natural broadening smooths the kinematic edges and reduces the
spectral difference between neutrino-mass hypotheses, although it does not
shift the edge reference positions.  For the transitions considered here,
this effect makes neutrino-mass measurements below ${\cal O}(1\,{\rm eV})$
particularly challenging, because the mass-induced edge displacement scales
as $m_\nu^2$ and rapidly becomes much smaller than the edge-smearing scale set
by the natural linewidth.  One possible direction is to identify alternative
transitions involving longer-lived atomic states and hence smaller natural
linewidths, while accounting for collisional broadening.
Another possibility is to revisit laser-induced fluorescence and Doppler
absorption spectroscopy, as proposed for tritium BSBD in
Ref.~\cite{cohen1987bound}, and apply Doppler-edge velocimetry
\cite{gentry1994edge,mckay1998modeling,mcgill1998comparison}. 

Thermal Doppler broadening poses a distinct challenge.  The distribution of
the parent-atom velocity projected along the detected photon direction
produces Gaussian broadening of the kinematic edges, without shifting their
reference positions.  One possibility -- albeit a highly speculative one --
would be to employ quantum-sensing technologies and use tagged tritium atoms
to accurately measure the velocities of parent atoms prior to their decay.
This information could in principle be used to reduce the parent-motion
contribution to the Doppler broadening on an event-by-event basis.  

Our analysis has focused on describing these physical effects and on
identifying the corresponding line-shape scales that can limit a
neutrino-mass measurement.  The sensitivity of a given experiment is beyond
the scope of this work and must be determined from a complete line-shape fit
including the setup-specific inputs and systematic effects.
 
An important advantage of our approach to direct neutrino mass measurement 
is that the fractional number of events in the neutrino-mass-sensitive 
region scales as $(m_\nu/Q)^2$. This is to be compared with the fractional 
number of useful events in the conventional CSBD-based approaches, 
which scales as $(m_\nu/Q)^3$. Thus, the gain factor is $Q/m_\nu$ in our 
case, which is a very large number, especially for light neutrinos.  
A disadvantage of our approach is that it imposes more stringent 
requirements on the acceptable temperature of the tritium source, as 
discussed above. 

In the limit of vanishing neutrino mass, the BSBD photon Doppler edge
determines the channel $Q$-value. For a fixed final atomic state,
$\Delta\omega=\omega_0 Q/M$ (where $M$ is the daughter-atom mass and
$\omega_0$ the unshifted transition frequency), and thus, 
$Q=M\Delta\omega/\omega_0$. For independent inputs, the relative uncertainty
is $\delta Q/Q=\sqrt{(\delta M/M)^2+
[\delta(\Delta\omega)/\Delta\omega]^2+
(\delta\omega_0/\omega_0)^2}$.
For the ${}^{3}\mathrm{He}$ triplet $1s3s\rightarrow1s2p$ red transition,
$\Delta\omega\simeq2.8~\mathrm{GHz}$. The AME2020 value
$\delta M/M=1.99\times10^{-11}$ \cite{AME2020II}, together with the
comparable or smaller fractional uncertainty on $\omega_0$, is negligible
relative to the Doppler-edge uncertainty in both benchmarks. Adopting the
demonstrated helium red-line precision
$\delta(\Delta\omega)=5~\mathrm{MHz}$ as an edge-measurement benchmark
\cite{Thomas_2020} gives
$\delta(\Delta\omega)/\Delta\omega\simeq1.8\times10^{-3}$ and therefore
$\delta Q\simeq33~\mathrm{eV}$. Prospectively, helium frequency-comb
spectroscopy indicates achievable relative optical-frequency precisions of
order $10^{-11}$ \cite{CancioPastor:2012helium}, corresponding to absolute
uncertainties from a few to a few tens of kilohertz near $706.5~\mathrm{nm}$.
If comparable absolute accuracy can be transferred to the BSBD Doppler-edge
extraction, the channel-$Q$ sensitivity would reach
$\mathcal{O}(100~\mathrm{meV})$, provided that natural-linewidth, thermal,
instrumental, calibration, and background effects are controlled at the same
frequency scale.

The direct neutrino measurement method proposed in this paper opens 
a new avenue for determining neutrino masses. Much work still has to be 
done, however, to establish whether it can constitute a viable 
alternative to existing methods 
and achieve sufficiently high sensitivity to neutrino masses.

\acknowledgments
{The authors are grateful to Oleg Chkvorets, Jose Crespo L\'opez-Urrutia, 
Guido Drexlin, J\"org Evers, Susanne Mertens, Daniel Murnick, Thomas Pfeifer, 
Hamish Robertson, Vera Sch\"{a}fer, Alexei Smirnov and 
Sergey Vasiliev for very useful and stimulating discussions.}

\appendix
\section{\label{sec:appA} 
2-body decay and the speed of ${\rm ^3He}$}
We use the following notation: $M_i\equiv M({\rm T})$ 
and $M_f\equiv M({\rm ^3He})$ are the atomic masses of tritium and 
${\rm ^3He}$, $m_k$ is mass of the $k$th neutrino mass eigenstate; 
$Q\equiv M_i-M_f$. 
For transitions to an excited atomic state of 
${\rm ^3He}$ with the excitation energy $E_{\rm exc}$, the quantity 
$M_f$ must be replaced by $M({\rm ^3He})+E_{\rm exc}$, and the $Q$-value 
of the process must be modified accordingly. 
In the rest frame of the parent tritium the 
velocity of the ${\rm ^3He}$ atom produced along with neutrino mass 
eigenstate $\nu_k$ is given by
\be
v_k=\frac{\left[(M_i^2-M_f^2)^2-2(M_i^2+M_f^2) m_k^2 
+m_k^4\right]^{1/2}}{M_i^2+M_f^2-m_k^2}\,. 
\label{eq:v_1}
\ee
Neglecting terms of the fourth and higher order in neutrino mass and 
taking into account that $Q^2/(M_i^2+M_f^2)\ll 1$, we arrive at 
eq.~(\ref{eq:v}), which is valid to an accuracy of about $10^{-11}$. 
If higher accuracy is needed, 
the exact formula (\ref{eq:v_1}) should be used. 

\section{\label{sec:appMixed} BSBD with a mixed atomic/molecular tritium 
source and background from CSBD}

We follow the notation of the main text: $\Gamma_1$ is the ordinary CSBD 
decay rate per tritium nucleus, and $\Gamma_2$ denotes here the 
effective BSBD production rate, which includes both singlet and triplet 
final-state configurations of
${}^3\mathrm{He}$ in the unresolved $1s3s$ states that yield red 
photons. The numerical benchmark uses
$\Gamma_2=(4/3)\Gamma_2^{\rm triplet}$.  The
controlled variable is the atomic fraction $k$, defined as the fraction of
tritium nuclei in atomic form.  For a $\beta$-activity $A$,
\be
  N_0=\frac{A}{\Gamma_1},\qquad
  N_a=kN_0,\qquad
  N_m=\frac{1}{2}(1-k)N_0 ,
  \label{eq:app_Na_Nm}
\ee
where $N_0$ is the total number of tritium nuclei, $N_a$ the number of
atomic tritium atoms, and $N_m$ the number of T$_2$ molecules. 
The observable is the red-line photoelectron yield from 
${\rm ^3He}^*(1s3s)$ 
singlet- and triplet state de-excitations.

The de-excitation of the $1s3s$ state may proceed through radiative or 
non-radiative channels. Non-radiative losses, photon transport, 
detector acceptance, optical transmission, collection efficiency, 
and photosensor quantum efficiency are absorbed into the overall 
efficiency factor $\varepsilon_{\rm tot}$. 
CSBD of T$_2$ can also populate the same unresolved 
states, directly or through dissociation, charge rearrangement, 
secondary excitation, or cascades.  This contribution is described 
by an effective $1s3s$-production yield $r$ per molecular CSBD 
decay.  A possible modification of the BSBD $1s3s$-production 
rate in molecular tritium is parameterized as $\Gamma_2^{\rm 
mol}=(1-\delta)\Gamma_2$.

For an exposure $t_i$ at fixed $k_i$, the detected photoelectron yield is
the photoelectron version of eq.~(\ref{eq:Nph1}); the corresponding
photon-yield equations, including the $\delta=0$ limit, have already been
given in eqs.~(\ref{eq:Nph1}) and (\ref{eq:Nph2}) and are not repeated: 
\be
N_{\rm pe}(k_i)=N_{\rm ph}(k_i)\,\varepsilon_{\rm tot}=
A t_i\varepsilon_{\rm tot}
\left[
\frac{\Gamma_2}{\Gamma_1}
+(1-k_i)
\left(r-\frac{\Gamma_2}{\Gamma_1}\delta\right)
\right].
\label{eq:app_photoelectron_yield_general}
\ee
The detected photoelectron yield, $N_{\rm pe}$, 
 is a linear function of $(1-k)$, with its slope depending in 
general on the value of $\delta$. For $\delta=0$ the slope 
simplifies to
$A\,t_i\,\varepsilon_{\rm tot}\,r$.
Therefore, in this limit, a scan of the atomic fraction $k$ 
provides two independent observables through the linear dependence of 
$N_{\rm pe}$ on $(1-k)$,
allowing the extraction of both the effective BSBD production rate
$\Gamma_2$ and the molecular CSBD yield $r$. This separation is
model-dependent: it assumes that atomic and molecular BSBD have the same
effective $1s3s$-production rate per tritium nucleus. If $\delta\neq0$, 
the coefficient of $(1-k)$ instead measures
$r_{\rm eff}=r-(\Gamma_2/\Gamma_1)\delta$. As discussed in the main text,
there are theoretical and phenomenological arguments suggesting that the
correction term $\Gamma_2/\Gamma_1 \delta$ is expected to be negligible,
such that $r_{\rm eff}\approx r$ in practice.

\begin{table}[t]
\centering
\small
\setlength{\tabcolsep}{4pt}
\label{tab:app_scan_inputs}
\begin{tabular}{@{}lcl@{\qquad}|lcl@{}}
\toprule
Quantity & Symbol & Value & Quantity & Symbol & Value \\
\midrule
Activity & $A$ & $10^9~\mathrm{s^{-1}}$ &
CSBD rate & $\Gamma_1$ & $1.786\times10^{-9}~\mathrm{s^{-1}}$ \\
Eff. BSBD rate & $\Gamma_2$ & $5.53\times10^{-13}~\mathrm{s^{-1}}$ &
CSBD yield & $r$ & $10^{-3}$ \\
Atomic scan & $k_i$ & $0,0.25,0.50,0.75$ &
Exposure/point & $t_i$ & $900~\mathrm{s}$ \\
Efficiency & $\varepsilon_{\rm tot}$ & $2.4\times10^{-3}$ &
Atomic-fraction error & $\sigma_k$ & $0.05$ \\
\bottomrule
\end{tabular}
\caption{\small Illustrative benchmark used to display the method.  
The CSBD yield $r$, scan range, exposure, and uncertainties are 
given for illustration purposes and should not be interpreted as an 
experimental forecast.  
}
\end{table}

The benchmark value $\varepsilon_{\rm tot}=2.4\times10^{-3}$ should therefore be 
interpreted as an effective detection efficiency that a $1s3s$ excitation
ultimately produces a detected photoelectron.  It is used only to
illustrate the analysis framework and should not be interpreted as the
performance of a specific detector design. 

The statistical treatment can be written in the standard pull-term form
\begin{equation}
  \chi^2(\xi,\phi,\eta_A,\eta_\varepsilon,\eta_{k,i}) =
  \sum_i
  \frac{\left[N_i-\mu_i(\xi,\phi,\eta_A,\eta_\varepsilon,\eta_{k,i})\right]^2}
       {\sigma_{{\rm stat},i}^2}
  +\eta_A^2+\eta_\varepsilon^2+\sum_i\eta_{k,i}^2 ,
  \label{eq:app_chi2}
\end{equation}
with
\begin{equation}
  \mu_i =
  (1+\eta_A\sigma_A)(1+\eta_\varepsilon\sigma_\varepsilon)
  \left[\xi+\phi\left(1-k_i-\eta_{k,i}\sigma_k\right)\right] .
  \label{eq:app_pull_model}
\end{equation}
The fit parameters $\xi$ and $\phi$ are the nuisance-free intercept
and molecular-fraction slope of the linearized yield model, both measured
in detected photoelectrons.  For the equal-exposure benchmark
$t_i=t$, comparison with eq.~(\ref{eq:app_photoelectron_yield_general})
gives 
\[
  \xi=A t\varepsilon_{\rm tot}\frac{\Gamma_2}{\Gamma_1},\qquad
  \phi=A t\varepsilon_{\rm tot}
  \left(r-\frac{\Gamma_2}{\Gamma_1}\delta\right).
\]
For unequal exposures, the same parametrization should be applied to
exposure-normalized yields or generalized to point-dependent
$\xi_i$ and $\phi_i$. 
Here $N_i$ is the detected photoelectron yield at scan point $i$, and
$\sigma_{{\rm stat},i}\simeq\sqrt{N_i}$ for large counts.  The nuisance
parameters $\eta_A$, $\eta_\varepsilon$, and $\eta_{k,i}$ have unit
Gaussian priors.  The benchmark uses fractional normalization widths
$\sigma_A=\sigma_\varepsilon=0.05$ and an absolute point-to-point
composition width $\sigma_k=0.05$.  A realistic analysis may instead split
the composition uncertainty into correlated calibration and uncorrelated
components.  For the reference case, the mean generated yields per 
measurement point are $\langle N_{ph}\rangle = 8.41\times10^8$ and
$\langle N_{\rm pe}\rangle = 2.02\times10^6$. The statistical uncertainty 
is only $\simeq0.07\%$, while the combined systematic uncertainty is
$\simeq9.3\%$ for the assumed inputs.  This illustrates the generic 
importance of source-composition and normalization systematics, but does 
not provide a quantitative sensitivity projection.

\begin{figure}[!ht]
\centering
\includegraphics[width=\textwidth]{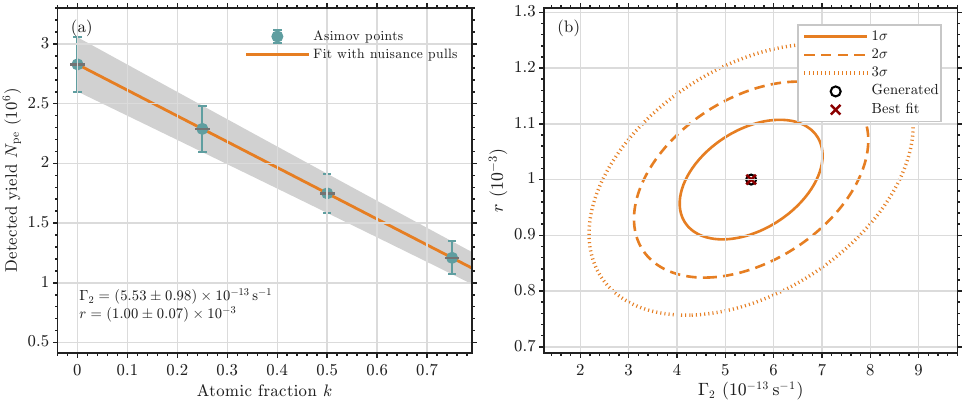}
\caption{\small Illustrative atomic-fraction scan for the benchmark 
configuration. Panel (a): detected photoelectron yield as a 
function of $k$, fitted with 
Eq.~\eqref{eq:app_photoelectron_yield_general} for $\delta=0$. 
The grey band shows the profiled nuisance-pull uncertainty. Panel 
(b): confidence regions in the $(\Gamma_2,r)$ plane obtained from 
$\Delta\chi^2=\chi^2-\chi^2_{\min}$. The contours are derived 
from the expected (unfluctuated) photoelectron yields for the 
benchmark values $\Gamma_2=5.53\times10^{-13}~\mathrm{s^{-1}}$ 
and $r=10^{-3}$.}
\label{fig:app_scan_summary}
\end{figure}

For the benchmark configuration considered here, fitting the 
expected (unfluctuated) yields gives 
$\Gamma_2=(5.53\pm0.98)\times10^{-13}~\mathrm{s^{-1}}$ and 
$r=(1.00\pm0.071)\times10^{-3}$. The quoted uncertainties are specific to the 
benchmark configuration considered here, including the assumed value of $r$, the 
choice of the four $k$ measurement points, and the adopted nuisance-parameter 
model. They are therefore intended to illustrate the fitting methodology rather 
than to represent a projected experimental sensitivity. The method is particularly 
useful because it separates the two
production mechanisms. The linear term in $(1-k)$ probes the molecular
CSBD-induced production of ${\rm ^3He}^*(1s3s)$, while the constant term
corresponds to the BSBD contribution in the reference model
($\delta=0$).}
A quantitative reach requires 
external experimental inputs, in particular the achievable range 
and stability of $k$, the optical collection efficiency, possible 
non-radiative quenching, the value of $r$, and non-BSBD optical 
backgrounds from walls, plasma processes, RF excitation, and 
delayed fluorescence.

Beyond the BSBD search itself, such a scan could provide an {\it in-situ}   
handle on the atomic and molecular fractions in a gaseous tritium 
source.  If calibrated against independent composition 
measurements, the $k$-dependent optical response could therefore 
become a useful source-diagnostics tool for future tritium-based 
neutrino-mass experimental programs.

\section{\label{sec:appStat} Statistical requirements for neutrino mass 
measurements}

Consider a ${\rm ^3He}$ emission line in the case of two neutrino 
species: one massless neutrino and one massive neutrino of mass $m_k$ 
(see Fig.~\ref{fig:box}). In the wings of the photon line profile 
($a_k<|\Delta|<a_0$), contributions from the massive neutrino 
are kinematically forbidden, and the spectrum of the emitted photons 
is entirely due to the massless neutrino. 
In the central region ($|\Delta|<a_k$), both massless 
and massive neutrinos contribute, with the weights $1-|U_{ek}|^2$ and 
$|U_{ek}|^2$, respectively. The heights of the lineshape plateaus in 
the wings and in the central region, 
denoted respectively by $F_w$ and $F_c$, are  
\be
F_w = \frac{1}{2a_0}\left(1-|U_{ek}|^2\right)\,,\qquad
F_c = \frac{1}{2a_0}\Big[1+|U_{ek}|^2\frac{a_0-a_k}{a_k}\Big]\;
\simeq\;\frac{1}{2a_0}\Big[1+|U_{ek}|^2\frac{\Delta v}{v_0}\Big], 
\label{eq:Fs}
\ee
The height of the steps at $\Delta=\pm a_k$ is therefore   
\be
\Delta F=F_c-F_w=\frac{1}{2a_0}
|U_{ek}|^2
\Big(1+\frac{\Delta v}{v_0}\Big).
\label{eq:steps}
\ee	

Let $N$ be the full number of the observed photon events corresponding 
to the atomic transition under discussion. The numbers of  
events corresponding to the wings and to the central region of the line 
profile, which we denote by $N_w$ and $N_c$, are 
proportional to the integrals of the photon spectrum over these regions, 
i.e.\ to the areas of the corresponding parts of the line profile 
shown in Fig.~\ref{fig:box}. From eq.~(\ref{eq:Fs}) we find
\be
N_w=N\frac{\Delta v}{v_0}\left(1-|U_{ek}|^2\right)\,,
\quad N_c=N\Big[1-\frac{\Delta v}{v_0}
\left(1-|U_{ek}|^2\right)\Big].
\label{eqNs}
\ee
To estimate the statistical significance of the neutrino mass 
determination, consider the number $N_c'$ of observed photons in the 
central region of the line 
profile, but restricted to 
the intervals in $\Delta$ of the same widths as the wings, i.e.\ in the 
ranges $2a_k-a_0<|\Delta|<a_k$:  
\be
N_c' \,=\,
\frac{\Delta v/v_0}{1-\Delta v/v_0}N_c \,\simeq \,
N\frac{\Delta v}{v_0}\Big[1+\frac{\Delta v}{v_0}|U_{ek}|^2\Big].
\label{eq:Ncprime}
\ee
A nonzero difference $N_c'-N_w \,\simeq\,
N\frac{\Delta v}{v_0}|U_{ek}|^2$ would signify nonvanishing 
neutrino mass. The statistical significance of observing $m_k\ne 0$ 
(the number of standard deviations) is then 
\be
Z~=~\frac{N_c'-N_w}{\sqrt{N_c'+N_w}}~\simeq
~\frac{\sqrt{N\frac{\Delta v}{v_0}}
|U_{ek}|^2}{\sqrt{2-|U_{ek}|^2}}.
\label{eq:appZ}
\ee
This yields eq.~(\ref{eq:N}) for the number of observed events required 
for significance $k\sigma$.


\bibliographystyle{JHEP}
\bibliography{bibliography}

\end{document}